\documentclass{article}
\usepackage{amsmath}
\usepackage{verbatim}
\usepackage[margin=1in]{geometry}
\makeatletter
\newcommand*{\rom}[1]{\expandafter\@slowromancap\romannumeral #1@}
\title{Moments and Cumulants of The Two-Stage Mann-Whitney Statistic}
\author{Dewei Zhong and John Kolassa}
\begin{document}
	\setlength{\baselineskip}{20pt}
	\maketitle
	\begin{abstract}
	This paper illustrates how to calculate the moments and cumulants of the two-stage Mann-Whitney statistic. These results may be used to calculate the asymptotic critical values of the two-stage Mann-Whitney test. In this paper, a large amount of deductions will be showed.
	\end{abstract}
	\section{Introduction to The Mann-Whitney Statistic and Two-Stage Test }
	The Mann-Whitney U test, is often used to test a difference in the responses of two groups.  $X_1,.....X_M$ are from control group,  $Y_1,......Y_N$ are from treatment group. The Mann-Whitney statistic is defined as\\
	\begin{equation}
		U=\sum\limits_{j=1}^{N}\sum\limits_{i=1}^{M}I_{ij},\quad\text{for}\quad I_{ij}=I(X_i<Y_j)=\begin{cases}
		1,&X_i<Y_j \\0,&X_i\geq Y_j
		\end{cases}.
	\end{equation}
    We select c as the critical value, If $U$ is larger than $c$, conclude that the two populations are different. This is one-stage test proposed by Mann and Whitney [1].
    \par
    Due to ethcal concerns and resource management, common designs allow for early stoppage in the presence of early and convincing proof. Spurrier and Hewett [2] provide a two-stage test based on the Mann-Whitney statistic. The two-stage test has two critical values. Denote these as $c_1$ and $c_2$. First, gather $m$ observations from control group and $n$ observations from treatment group.  Define
    \begin{equation}
    	U_1=\sum\limits_{j=1}^n\sum\limits_{i=1}^mI_{ij}.
    \end{equation}
    Calculate $U_1$, and if it is larger than or equal to $c_1$, we can stop trials early to declare the treatment group is superior to control group. If $U_1$ is less than the first critical value $c_1$, gather $M-m$ obervations from control group and $N-n$ observations from treatment group, where$\quad m\leq M,n\leq N$. Define 
    \begin{equation}
    	U_2=\sum\limits_{j=1}^N\sum\limits_{i=1}^MI_{ij}.
    \end{equation}
    Then calculate $U_2$. If it is larger than or equal the second $c_2$, we claim the treated is superior to the controls. 
    \par The hard part of this test is to find two critical values. The critical values of Mann-Whitney statistic in one dimension as well as critical values of one stage test can be easily calculated. We focus the critical values for two stage test. Due to the complexity of the mass function for two dimensional Mann-Whitney statistics, the complexity to get exact critical values is computatively intensive. Therefore, the asymptotic critical values within tolarable error are decent substitutions. The Cornish Fisher expansion provides the method to obtain critical values [3]. In order to use the Cornish Fisher expansion, the cumulants are necessary. So next, we will find the moments for two stage test statistics $U_1$ and $U_2$ in section 2 and get the cumulants in section 3.
	\section{Moments of The Two Dimensional Mann-Whitney Statistic}
	First, we assume $X_1\cdots X_M$ and $Y_1\cdots Y_M$ are jointly independent and identically distributed. Under the null hypthesis, all of the observations belong to the same population. In the first subsection, we give some hyphenated transition. In the second, third, fourth and fifth subsection, we find the first, second, third and fourth moments repectively under both the general case and the null hypothesis.
	\subsection{Probability Definition}
    Some transition values will be used in the following deductions. We give them first. The $I_{ij}$ is the same as that in (1).\\ Under the null hypothesis:
    \begin{equation}
    \begin{split}
    E(I_{ij})=\frac{1}{2} \qquad
    E(I_{ij}I_{kj})=\frac{1}{3} \qquad
    E(I_{ij}I_{il})=\frac{1}{3} \qquad
    E(I_{ij}I_{kl})=\frac{1}{4} \\ 
    \qquad E(I_{ij}I_{il}I_{it})=\frac{1}{4}\qquad
    E(I_{ij}I_{kj}I_{sj})=\frac{1}{4}  \qquad
    E(I_{ij}I_{kj}I_{kl})=\frac{5}{24} \\
    E(I_{ij}I_{kj}I_{sj}I_{st})=\frac{3}{20}\qquad 
    E(I_{ij}I_{kj}I_{it}I_{st})=\frac{2}{15}\qquad
    E(I_{ij}I_{kj}I_{sj}I_{pj})=\frac{1}{5}\\
    E(I_{ij}I_{kj}I_{iq}I_{it})=\frac{3}{20}\qquad
    E(I_{ij}I_{kj}I_{il}I_{kt})=\frac{1}{6}\qquad
    E(I_{ij}I_{kj}I_{iq}I_{kl})=\frac{2}{15}\\
    E(I_{ij}I_{il}I_{kj})=\frac{5}{24}\qquad
    E(I_{ij}I_{il}I_{iq}I_{st})=\frac{1}{8}\qquad
    E(I_{ij}I_{kj}I_{pj}I_{st})=\frac{1}{8}\\
    E(I_{ij}I_{il}I_{pq}I_{pt})=\frac{1}{9}\qquad
    E(I_{ij}I_{kj}I_{pq}I_{sq})=\frac{1}{9}\qquad
    E(I_{ij}I_{il}I_{pq}I_{sq})=\frac{1}{9}\\
    E(I_{ij}I_{kj}I_{il}I_{st})=\frac{1}{9}\qquad
    E(I_{ij}I_{kj}I_{pq}I_{st})=\frac{1}{12}\qquad
    E(I_{ij}I_{kl}I_{st}I_{sq})=\frac{1}{12}\\
    E(I_{ij}I_{kl}I_{st}I_{pq})=\frac{1}{16}\qquad
    E(I_{ij}I_{il}I_{it}I_{iq})=\frac{1}{5}\qquad 
    E(I_{ij}I_{kj}I_{iq}I_{st})=\frac{5}{48}.\end{split}
    \end{equation}
	 In the general case, define the probabilities $\pi_i$ below:\\
	 \begin{equation}
	 \begin{split}
	 E(I_{ij})=\pi_0 \qquad
	 E(I_{ij}I_{kj})=\pi_1 \qquad
	 E(I_{ij}I_{kj}I_{sj})=\pi_2  \qquad
	 E(I_{ij}I_{kj}I_{sj}I_{st})=\pi_3\\
	 E(I_{ij}I_{kj}I_{kl})=\pi_4 \qquad
	 E(I_{ij}I_{kj}I_{it}I_{st})=\pi_5\qquad
	 E(I_{ij}I_{kj}I_{sj}I_{pj})=\pi_6\\
	 E(I_{ij}I_{kj}I_{iq}I_{it})=\pi_7\qquad
	 E(I_{ij}I_{kj}I_{il}I_{kl})=\pi_8\qquad
	 E(I_{ij}I_{il})=\pi_9\\
	 E(I_{ij}I_{kj}I_{il}I_{kt})=\pi_{10}\qquad
	 E(I_{ij}I_{il}I_{it})=\pi_{12}\qquad
	 E(I_{ij}I_{il}I_{it}I_{iq})=\pi_{13}.\\ \\
	 \end{split}
	 \end{equation} \\
	 Here, $I_{ij}I_{il}I_{kj}=1\quad$ implied that all of $X_i<Y_j, \quad X_i<Y_l,\quad X_k<Y_j$ hold. The equaiton $\quad I_{ij}I_{il}I_{kj}=0\quad$ implies at least one of them does not hold. Here we intend $i,k,s,p\quad$in the same expression are pairwise unequal, as are $j,l,q,t$. For example, in $I_{ij}I_{kl}I_{pq}I_{st}$, $i\neq k,i\neq p, i\neq s, k\neq p,k\neq s, p\neq s$ and $j\neq l,j\neq t,j\neq q,l\neq q,l\neq t,q\neq t$.
	\subsection{First Moments}
	In genral, by (5)
	\begin{align}
	&E(U_1)=\sum\limits_{j=1}^{n}\sum\limits_{i=1}^{m}E(I_{ij})=mn\pi_0, \\ &E(U_2)=MN\pi_0. 
	\end{align}
 Under the null hypothesis, by (4)
 \begin{align}
 	&E(U_1)=\sum\limits_{j=1}^{n}\sum\limits_{i=1}^{m}E(I_{ij})=\frac{mn}{2},  \\& E(U_2)=\frac{MN}{2}. 
 \end{align}
	\subsection{Second Moments}
	\begin{align}
	\begin{split}
	U_1^2&=(\sum\limits_{i=1}^{m}\sum\limits_{j=1}^{n}I_{ij} )^2\\ &=\sum\limits_{i=1}^{m}\sum\limits_{j=1}^{n}I_{ij}^2+\sum\limits_{i=1}^{m}\sum\limits_{j=1}^{n}\sum\limits_{k=1,k\neq i}^{m} I_{ij}I_{kj}+\sum\limits_{j=1}^{n}\sum\limits_{i=1}^{m}\sum\limits_{l=1,l\neq j}^{n} I_{ij}I_{il}+\\ &\quad+\sum\limits_{j=1}^{n}\sum\limits_{i=1}^{m}\sum\limits_{l=1,l\neq j}^{n}\sum\limits_{k=1,k\neq i}^{m} I_{ij}I_{kl}.
	\end{split}
	\end{align}
	Substituting the probability values in (10) by (5)
	\begin{align}
	E(U_1^2)&=mn\pi_0+m(m-1)n\pi_1+mn(n-1)\pi_9+m(m-1)n(n-1)\pi_0^2.
	\end{align}
	By the same reasoning, 
	\begin{align}	E(U_2^2)&=MN\pi_0+M(M-1)N\pi_1+MN(N-1)\pi_9+M(M-1)N(N-1)\pi_0^2.
	\end{align}
	While calculating mixed moments, conditional expectations are usually utilized. 
	\begin{align}
	\begin{split}
	E(U_1U_2)&=E(E(U_1U_2|U_2))\\&=E(U_2E(U_1|U_2))\\&=E(U_2\frac{mn}{MN}U_2)\\&=\frac{mn}{MN}E(U_2^2).
	\end{split}
	\end{align}
	Thus, by (12) and (13)
	\begin{align}
	E(U_1U_2)&=mn\pi_0+mn(M-1)\pi_1+mn(N-1)\pi_9+mn(M-1)(N-1)\pi_0^2.
	\end{align}
	Under the null hypothesis, by (4) and (14):
	\begin{align}
	\begin{split}
	E(U_1^2)&=\frac{1}{2}mn+\frac{1}{3}m(m-1)n+\frac{1}{3}mn(n-1)+\frac{1}{4}m(m-1)n(n-1)\\&=\frac{m^2n^2}{4}+\frac{m^2n}{12}+\frac{mn^2}{12}+\frac{mn}{12},
	\end{split}
	\end{align}
	By the same reasoning, 
	\begin{align}
	E(U_2^2)&=\frac{M^2N^2}{4}+\frac{M^2N}{12}+\frac{MN^2}{12}+\frac{MN}{12},
\end{align}
\begin{align} 
\begin{split}
	 E(U_1U_2)&=E(E(U_1U_2|U_2))\\&=E(U_2E(U_1|U_2))\\&=E(U_2\frac{mn}{MN}U_2)\\&=\frac{mn}{MN}E(U_2^2),
	 \end{split}
\end{align}
\begin{align}
	E(U_1U_2)&=\frac{mnMN}{4}+\frac{mnM}{12}+\frac{mnN}{12}+\frac{mn}{12}.
	\end{align}
	\subsection{Third Moments}
	We extend calculations for lower orders to find it.
	\begin{align}
	\begin{split}
    E(U_1^3)&=\sum\limits_{j=1}^n\sum\limits_{i=1}^mI_{ij}+3\sum\limits_{j=1}^n\sum\limits_{i=1}^m\sum\limits_{k=1,k\neq i}^mI_{ij}I_{ij}I_{kj}+3\sum\limits_{i=1}^m\sum\limits_{j=1}^n\sum\limits_{l=1,l\neq j}^nI_{ij}I_{ij}I_{il}\\&+3\sum\limits_{j=1}^n\sum\limits_{i=1}^m\sum\limits_{k=1,k\neq i}^m\sum\limits_{l=1,l\neq j}^nI_{ij}I_{ij}I_{kl}+6\sum\limits_{j=1}^n\sum\limits_{i=1}^m\sum\limits_{k=1,k\neq i}^m\sum\limits_{l=1,l\neq j}^nI_{ij}I_{il}I_{kj}\\&+\sum\limits_{j=1}^n\sum\limits_{i=1}^m\sum\limits_{k=1,k\neq i}^m\sum\limits_{s=1,s\neq i,s\neq k}^mI_{ij}I_{kj}I_{sj}+\sum\limits_{j=1}^n\sum\limits_{i=1}^m\sum\limits_{l=1,l\neq j}^n\sum\limits_{t=1,t\neq j, t\neq l}^nI_{ij}I_{il}I_{it}\\&+3\sum\limits_{j=1}^n\sum\limits_{i=1}^m\sum\limits_{k=1,k\neq i}^m\sum\limits_{l=1,l\neq j}^n\sum\limits_{s=1,s\neq i,s\neq k}^mI_{ij}I_{kl}I_{sj}\\&+3\sum\limits_{j=1}^n\sum\limits_{i=1}^m\sum\limits_{k=1,k\neq i}^m\sum\limits_{l=1,l\neq j}^n\sum\limits_{t=1,t\neq j, t\neq l}^nI_{ij}I_{kl}I_{it}\\&+\sum\limits_{j=1}^n\sum\limits_{i=1}^m\sum\limits_{k=1,k\neq i}^m\sum\limits_{l=1,l\neq j}^n\sum\limits_{s=1,s\neq i,s\neq k}\sum\limits_{t=1,t\neq j, t\neq l}^nI_{ij}I_{kl}I_{st}.
    \end{split}
    \end{align}
    Substituting (5) in (19),
	\begin{align}
	\begin{split}
	E(U_1^3)&=mn\pi_0+3m(m-1)n\pi_1+3mn(n-1)\pi_9+3m(m-1)n(n-1)\pi_0^2\\&+6m(m-1)n(n-1)\pi_4+m(m-1)(m-2)n\pi_2+mn(n-1)(n-2)\pi_{12}\\&+3m(m-1)(m-2)n(n-1)\pi_0\pi_1+3m(m-1)n(n-1)(n-2)\pi_0\pi_9\\&+m(m-1)(m-2)n(n-1)(n-2)\pi_0^3.
	\end{split}
	\end{align}
	Simlifying,
	\begin{align}
	\begin{split}
		E(U_1^3)&=mn\pi_0+3m(m-1)n\pi_1+3mn(n-1)\pi_9+3m(m-1)n(n-1)\pi_0^2\\&+6m(m-1)n(n-1)\pi_4+m(m-1)(m-2)n\pi_2+mn(n-1)(n-2)\pi_{12}\\&+3m(m-1)(m-2)n(n-1)\pi_0\pi_1+3m(m-1)n(n-1)(n-2)\pi_0\pi_9\\&+m(m-1)(m-2)n(n-1)(n-2)\pi_0^3.
	\end{split}
	\end{align}
	By the same reasoning, 
	\begin{align}
	\begin{split}
	E(U_2^3)&=MN\pi_0+3M(M-1)N\pi_1+3MN(N-1)\pi_9+3M(M-1)N(N-1)\pi_0^2\\&+6M(M-1)N(N-1)\pi_4+M(M-1)(M-2)N\pi_2+MN(N-1)(N-2)\pi_{12}\\&+3M(M-1)(M-2)N(N-1)\pi_0\pi_1+3M(M-1)N(N-1)(N-2)\pi_0\pi_9\\&+M(M-1)(M-2)N(N-1)(N-2)\pi_0^3.
	\end{split}
	\end{align}
	Next, the conditional expectations are used to get $E(U_1U_2^2)$ and $E(U_1^2U_2)$.
	\begin{align}
	E(U_1U_2^2)=E[U_2^2E(U_1|U_2)]=E(U_2^2\frac{mn}{MN}U_2)=\frac{mn}{MN}E(U_2^3).
	\end{align}
	Substituting (22) in (23),
	\begin{align}
	\begin{split}
	E(U_1U_2^2)&=mn\pi_0+3mn(M-1)\pi_1+3mn(N-1)\pi_9+3mn(M-1)(N-1)\pi_0^2\\&+6mn(M-1)(N-1)\pi_4+mn(M-1)(M-2)\pi_2+mn(N-1)(N-2)\pi_{12}\\&+3mn(M-1)(M-2)(N-1)\pi_0\pi_1+3mn(M-1)(N-1)(N-2)\pi_0\pi_9\\&+mn(M-1)(M-2)(N-1)(N-2)\pi_0^3.
	\end{split}
	\end{align}
	\begin{align}
	\begin{split}
	E(U_1^2U_2)=&E(\sum\limits_{i=1}^m\sum\limits_{j=1}^{n}\sum\limits_{k=1}^m\sum\limits_{l=1}^nI_{ij}I_{kl}U_2)\\=&E(\sum\limits_{i=1}^m\sum\limits_{j=1}^{n}\sum\limits_{k=1, k\neq i}^m\sum\limits_{l=1, l\neq j}^nI_{ij}I_{kl}U_2+\sum\limits_{i=1}^m\sum\limits_{j=1}^n\sum\limits_{k=1, k\neq i}^mI_{ij}I_{kj}U_2+\sum\limits_{i=1}^m\sum\limits_{j=1}^n\sum\limits_{l=1,l \neq j}^nI_{ij}I_{il}U_2\\&+\sum\limits_{i=1}^m\sum\limits_{j=1}^nI_{ij}U_2)\\=&E(\frac{m(m-1)n(n-1)}{M(M-1)N(N-1)}\sum\limits_{i=1}^M\sum\limits_{j=1}^{N}\sum\limits_{k=1,k \neq i}^M\sum\limits_{l=1, l\neq j}^NI_{ij}I_{kl}U_2)\\&+E(\frac{m(m-1)n}{M(M-1)N}\sum\limits_{i=1}^M\sum\limits_{j=1}^N\sum\limits_{k=1, k\neq i}^MI_{ij}I_{kj}U_2)\\&+E(\frac{mn(n-1)}{MN(N-1)}\sum\limits_{i=1}^M\sum\limits_{j=1}^N\sum\limits_{l=1,l \neq j}^NI_{ij}I_{il}U_2)+E(\frac{mn}{MN}\sum\limits_{i=1}^M\sum\limits_{j=1}^NI_{ij}U_2)
	\\=&E(\frac{m(m-1)n(n-1)}{M(M-1)N(N-1)}\sum\limits_{i=1}^M\sum\limits_{j=1}^{N}\sum\limits_{k=1}^M\sum\limits_{l=1}^NI_{ij}I_{kl}U_2)\\&+E((\frac{m(m-1)n}{M(M-1)N}-\frac{m(m-1)n(n-1)}{M(M-1)N(N-1)})\sum\limits_{i=1}^M\sum\limits_{j=1}^N\sum\limits_{k=1, k\neq i}^MI_{ij}I_{kj}U_2)\\&+E((\frac{mn(n-1)}{MN(N-1)}-\frac{m(m-1)n(n-1)}{M(M-1)N(N-1)})\sum\limits_{i=1}^M\sum\limits_{j=1}^N\sum\limits_{l=1,l \neq j}^NI_{ij}I_{il}U_2)\\&+E((\frac{mn}{MN}-\frac{m(m-1)n(n-1)}{M(M-1)N(N-1)})\sum\limits_{i=1}^M\sum\limits_{j=1}^NI_{ij}U_2)\\=&\frac{m(m-1)n(n-1)}{M(M-1)N(N-1)}E(U_2^3)+(\frac{mn}{MN}-\frac{m(m-1)n(n-1)}{M(M-1)N(N-1)})E(U_2^2)\\&+(\frac{m(m-1)n}{M(M-1)N}-\frac{m(m-1)n(n-1)}{M(M-1)N(N-1)})E(\sum\limits_{i=1}^M\sum\limits_{j=1}^N\sum\limits_{k=1, k\neq i}^MI_{ij}I_{kj}U_2)\\&+(\frac{mn(n-1)}{MN(N-1)}-\frac{m(m-1)n(n-1)}{M(M-1)N(N-1)})E(\sum\limits_{i=1}^M\sum\limits_{j=1}^N\sum\limits_{l=1,l \neq j}^NI_{ij}I_{il}U_2).
	\end{split}
	\end{align}
	We can find
	\begin{align*}
	E(\sum\limits_{i=1}^M\sum\limits_{j=1}^N\sum\limits_{k=1, k\neq i}^MI_{ij}I_{kj}U_2)=&E(\sum\limits_{i=1}^M\sum\limits_{j=1}^N\sum\limits_{k=1, k\neq i}^M\sum\limits_{l=1, l\neq j}^NI_{ij}I_{kj}I_{il}+\sum\limits_{i=1}^M\sum\limits_{j=1}^N\sum\limits_{k=1, k\neq i}^MI_{ij}I_{kj}I_{kl}\\&+\sum\limits_{i=1}^M\sum\limits_{j=1}^N\sum\limits_{k=1, k\neq i}^M\sum\limits_{s=1,s\neq i,s\neq k}^M\sum\limits_{l=1, l\neq j}^NI_{ij}I_{kj}I_{sl}\\&+
	\sum\limits_{i=1}^M\sum\limits_{j=1}^N\sum\limits_{k=1, k\neq i}^M\sum\limits_{s=1,s\neq i,s\neq k}^MI_{ij}I_{kj}I_{sj}+\sum\limits_{i=1}^M\sum\limits_{j=1}^N\sum\limits_{k=1, k\neq i}^M(I_{ij}I_{kj}I_{ij}+I_{ij}I_{kj}I_{kj}))\\=&\pi_0\pi_1M(M-1)(M-2)N(N-1)+2\pi_4M(M-1)N(N-1)\\&+\pi_2M(M-1)(M-2)N+2\pi_1M(M-1)N.
	\end{align*}
	By the same reasoning, 
	\begin{align*}
	E(\sum\limits_{i=1}^M\sum\limits_{j=1}^N\sum\limits_{l=1, l\neq j}^NI_{ij}I_{il}U_2)=&\pi_0\pi_9M(M-1)N(N-1)(N-2)+2\pi_4M(M-1)N(N-1)\\&+\pi_{12}MN(N-1)(N-2)+2\pi_9MN(N-1).
	\end{align*}
	$E(U_1^3)$ and $E(U_1^2)$ have been found in (21) and (11). Due to the complication of simplifying, we stop here.\\ 
	\emph{Under the null hypothesis}, we can just use the values in (4) to substitute the $\pi_0 \cdots \pi_{13}$ in (21), (22), (24) and (25) . And we will show another method to get it. \\The distribution of $U_1$ is symmetric about 0 under the null hypothesis. Hence
	\begin{align*}
	E[U_1-E(U_1)]^3=0, \qquad \text{and so}
	\end{align*}
	\begin{align}
	E(U_1^3)=[E(U_1)]^3-3E(U_1)[E(U_1)]^2+3E(U_1^2)E(U_1).
	\end{align}
	Substituting (8) and (15), then do simplifying, (27) is got. The same way for $U_2$ to find (28) and the same conditional expectations to (29). 
	\begin{align}
	&E(U_1^3)=\frac{m^3n^3}{8}+\frac{m^3n^2}{8}+\frac{m^2n^3}{8}+\frac{m^2n^2}{8},
	 \\&E(U_2^3)=\frac{M^3N^3}{8}+\frac{M^3N^2}{8}+\frac{M^2N^3}{8}+\frac{M^2N^2}{8},
	\\&E(U_1U_2^2)=\frac{mnM^2N^2}{8}+\frac{mnM^2N}{8}+\frac{mnMN^2}{8}+\frac{mnMN}{8}.
	\end{align}
	There is a property under the null hypothesis,
	\begin{align*}
	E(U_2|U_1)=\frac{M+N+1}{m+n+1}U_1+\frac{1}{2}[\frac{(M-m)n(n+1)+(N-n)m(m+1)}{m+n+1}+(M-m)(N-n)].
	\end{align*}
	And thus,
	\begin{align}
	\begin{split}
	E(U_1^2U_2)&=E[U_1^2E(U_2|U_1)]\\&=E[U_1^2(\frac{M+N+1}{m+n+1}U_1+\frac{1}{2}[\frac{(M-m)n(n+1)+(N-n)m(m+1)}{m+n+1}+(M-m)(N-n)])]\\&=\frac{M+N+1}{m+n+1}E(U_1^3)+\frac{1}{2}[\frac{(M-m)n(n+1)+(N-n)m(m+1)}{m+n+1}+(M-m)(N-n)]E(U_1^2)\\&=\frac{M+N+1}{m+n+1}\frac{m^2n^2}{8}(mn+m+n+1)\\&\quad+\frac{mn}{24}\frac{(M-m)n(n+1)+(N-n)m(m+1)}{m+n+1}(3mn+m+n+1)\\&\quad+\frac{1}{12}(M-m)(N-n))(3m^2n^2+m^n+mn^2+mn).
	\end{split}
	\end{align}
	The result is the same as (25) under the null hypothesis.
	\subsection{Fourth Moments}
	We extend calculations for lower order moments to find it,
	\begin{align*}
	U_1^4&=(\sum\limits_{j=1}^n\sum\limits_{i=1}^{m}I_{ij}I_{ij}I_{ij}I_{ij})^4\\&=\sum\limits_{j=1}^n\sum\limits_{i=1}^{m}I_{ij}I_{ij}I_{ij}I_{ij}+3\sum\limits_{j=1}^n\sum\limits_{i=1}^m\sum\limits_{k=1,k\neq i}^{m}I_{ij}I_{ij}I_{kj}I_{kj}+3\sum\limits_{j=1}^n\sum\limits_{i=1}^m\sum\limits_{l=1,l\neq j}^{n}I_{ij}I_{ij}I_{il}I_{il}\\&\qquad+4\sum\limits_{j=1}^n\sum\limits_{i=1}^m\sum\limits_{k=1,k\neq i}^{m}I_{ij}I_{ij}I_{ij}I_{kj}+4\sum\limits_{j=1}^n\sum\limits_{i=1}^m\sum\limits_{l=1,l\neq j}^{n}I_{ij}I_{ij}I_{ij}I_{il}+6\sum\limits_{j=1}^n\sum\limits_{i=1}^m\sum\limits_{l=1,l\neq j}^{n}\sum\limits_{t=1,t\neq j,t\neq l}^{n}I_{ij}I_{ij}I_{il}I_{it}\\&\qquad+ 6\sum\limits_{j=1}^n\sum\limits_{i=1}^m\sum\limits_{k=1,k\neq i}^{m}\sum\limits_{s=1,s\neq i,s\neq k}^{m}I_{ij}I_{ij}I_{kj}I_{sj}+12\sum\limits_{j=1}^n\sum\limits_{i=1}^m\sum\limits_{k=1,k\neq i}^{m}\sum\limits_{l=1,l\neq j}^{n}I_{ij}I_{ij}I_{kj}I_{kl}\\&\qquad+12\sum\limits_{j=1}^n\sum\limits_{i=1}^m\sum\limits_{k=1,k\neq i}^{m}\sum\limits_{l=1,l\neq j}^{n}I_{ij}I_{ij}I_{kj}I_{il}+12\sum\limits_{j=1}^n\sum\limits_{i=1}^m\sum\limits_{k=1,k\neq i}^{m}\sum\limits_{l=1,l\neq j}^{n}I_{ij}I_{ij}I_{kl}I_{il}\\&\qquad+6\sum\limits_{j=1}^n\sum\limits_{i=1}^m\sum\limits_{k=1,k\neq i}^{m}\sum\limits_{l=1,l\neq j}^{n}I_{ij}I_{il}I_{kj}I_{kl}+4\sum\limits_{j=1}^n\sum\limits_{i=1}^m\sum\limits_{k=1,k\neq i}^{m}\sum\limits_{l=1,l\neq j}^{n}I_{ij}I_{ij}I_{ij}I_{kl}\\&\qquad+3\sum\limits_{j=1}^n\sum\limits_{i=1}^m\sum\limits_{k=1,k\neq i}^{m}\sum\limits_{l=1,l\neq j}^{n}I_{ij}I_{ij}I_{kl}I_{kl}+6\sum\limits_{j=1}^n\sum\limits_{i=1}^m\sum\limits_{k=1,k\neq i}^{m}\sum\limits_{l=1,l\neq j}^{n}\sum\limits_{t=1,t\neq j,t\neq l}^{n}I_{ij}I_{ij}I_{kl}I_{kt}\\&\qquad +6\sum\limits_{j=1}^n\sum\limits_{i=1}^m\sum\limits_{k=1,k\neq i}^{m}\sum\limits_{l=1,l\neq j}^{n}\sum\limits_{s=1,s\neq i,s\neq k}^{m}I_{ij}I_{ij}I_{kl}I_{sl}+12\sum\limits_{j=1}^n\sum\limits_{i=1}^m\sum\limits_{k=1,k\neq i}^{m}\sum\limits_{l=1,l\neq j}^{n}\sum\limits_{t=1,t\neq j,t\neq l}^{n}I_{ij}I_{il}I_{it}I_{kj}\\&\qquad +12\sum\limits_{j=1}^n\sum\limits_{i=1}^m\sum\limits_{k=1,k\neq i}^{m}\sum\limits_{l=1,l\neq j}^{n}\sum\limits_{s=1,s\neq i,s\neq k}^{m}I_{ij}I_{sj}I_{il}I_{kj}+12\sum\limits_{j=1}^n\sum\limits_{i=1}^m\sum\limits_{k=1,k\neq i}^{m}\sum\limits_{l=1,l\neq j}^{n}\sum\limits_{t=1,t\neq j,t\neq l}^{n}I_{ij}I_{ij}I_{il}I_{kt}\\&\qquad +12\sum\limits_{j=1}^n\sum\limits_{i=1}^m\sum\limits_{k=1,k\neq i}^{m}\sum\limits_{l=1,l\neq j}^{n}\sum\limits_{s=1,s\neq i,s\neq k}^{m}I_{ij}I_{kj}I_{il}I_{sl}+12\sum\limits_{j=1}^n\sum\limits_{i=1}^m\sum\limits_{k=1,k\neq i}^{m}\sum\limits_{l=1,l\neq j}^{n}\sum\limits_{t=1,t\neq j,t\neq l}^{n}I_{ij}I_{kj}I_{il}I_{kt}\\&\quad+12\sum\limits_{j=1}^n\sum\limits_{i=1}^m\sum\limits_{k=1,k\neq i}^{m}\sum\limits_{l=1,l\neq j}^{n}\sum\limits_{s=1,s\neq i,s\neq k}^{m}I_{ij}I_{ij}I_{kj}I_{sl}+\sum\limits_{j=1}^n\sum\limits_{i=1}^m\sum\limits_{k=1,k\neq i}^{m}\sum\limits_{s=1,s\neq i,s\neq k}^{m}\sum\limits_{p=1,p\neq i,p\neq k,p\neq s}^{m}I_{ij}I_{kj}I_{sj}I_{pj}\\&\qquad+\sum\limits_{i=1}^m\sum\limits_{j=1}^n\sum\limits_{l=1,l\neq j}^{n}\sum\limits_{t=1,t\neq j,t\neq l}^{n}\sum\limits_{q=1,q\neq j,q\neq l,q\neq t}^{n}I_{ij}I_{iq}I_{it}I_{il}\\&\qquad +4\sum\limits_{i=1}^m\sum\limits_{k=1,k\neq i}^m\sum\limits_{j=1}^n\sum\limits_{l=1,l\neq j}^{n}\sum\limits_{t=1,t\neq j,t\neq l}^{n}\sum\limits_{q=1,q\neq j,q\neq l,q\neq t}^{n}I_{ij}I_{iq}I_{it}I_{kl}\\& \qquad +4\sum\limits_{j=1}^n\sum\limits_{l=1,l\neq j}^n\sum\limits_{i=1}^m\sum\limits_{k=1,k\neq i}^{m}\sum\limits_{s=1,s\neq i,s\neq k}^{m}\sum\limits_{p=1,p\neq i,p\neq k,p\neq s}^{m}I_{ij}I_{kj}I_{sj}I_{pl}\\&\qquad +3\sum\limits_{i=1}^m\sum\limits_{k=1,k\neq i}^m\sum\limits_{j=1}^n\sum\limits_{l=1,l\neq j}^{n}\sum\limits_{t=1,t\neq j,t\neq l}^{n}\sum\limits_{q=1,q\neq j,q\neq l,q\neq t}^{n}I_{ij}I_{il}I_{kt}I_{kq}\\ 
	\end{align*}
	\begin{align}
	\begin{split}
	&\qquad +3\sum\limits_{j=1}^n\sum\limits_{l=1,l\neq j}^n\sum\limits_{i=1}^m\sum\limits_{k=1,k\neq i}^{m}\sum\limits_{s=1,s\neq i,s\neq k}^{m}\sum\limits_{p=1,p\neq i,p\neq k,p\neq s}^{m}I_{ij}I_{kj}I_{sl}I_{pl}\\&\qquad+6\sum\limits_{j=1}^n\sum\limits_{i=1}^m\sum\limits_{k=1,k\neq i}^{m}\sum\limits_{l=1,l\neq j}^{n}\sum\limits_{s=1,s\neq i,s\neq k}^{m}\sum\limits_{t=1,t\neq j,t\neq l}^{n}I_{ij}I_{il}I_{kt}I_{st}\\&\qquad +6\sum\limits_{j=1}^n\sum\limits_{i=1}^m\sum\limits_{k=1,k\neq i}^{m}\sum\limits_{l=1,l\neq j}^{n}\sum\limits_{t=1,t\neq j,t\neq l}^{n}\sum\limits_{s=1,s\neq i,s\neq k}^mI_{ij}I_{ij}I_{kl}I_{st}\\&\qquad +24\sum\limits_{j=1}^n\sum\limits_{i=1}^m\sum\limits_{k=1,k\neq i}^{m}\sum\limits_{l=1,l\neq j}^{n}\sum\limits_{t=1,t\neq j,t\neq l}^{n}\sum\limits_{s=1,s\neq i,s\neq k}^mI_{ij}I_{kj}I_{il}I_{st}\\&\qquad +6\sum\limits_{j=1}^n\sum\limits_{i=1}^m\sum\limits_{k=1,k\neq i}^{m}\sum\limits_{l=1,l\neq j}^{n}\sum\limits_{t=1,t\neq j,t\neq l}^{n}\sum\limits_{s=1,s\neq i,s\neq k}^m\sum\limits_{p=1,p\neq i,p\neq k,p\neq s}^mI_{ij}I_{kj}I_{pl}I_{st}\\&\qquad +6\sum\limits_{j=1}^n\sum\limits_{i=1}^m\sum\limits_{k=1,k\neq i}^{m}\sum\limits_{l=1,l\neq j}^{n}\sum\limits_{t=1,t\neq j,t\neq l}^{n}\sum\limits_{s=1,s\neq i,s\neq k}^m\sum\limits_{p=1,p\neq i,p\neq k,p\neq s}^mI_{ij}I_{il}I_{kq}I_{st}\\&\qquad +\sum\limits_{j=1}^n\sum\limits_{i=1}^m\sum\limits_{k=1,k\neq i}^{m}\sum\limits_{l=1,l\neq j}^{n}\sum\limits_{t=1,t\neq j,t\neq l}^{n}\sum\limits_{s=1,s\neq i,s\neq k}^m\sum\limits_{p=1,p\neq i,p\neq k,p\neq s}^m\sum\limits_{q=1,q\neq j,q\neq l,q\neq t}^nI_{ij}I_{kl}I_{pq}I_{st},
	\end{split}
	\end{align}
	Substituting the values in (5),
	\begin{align}
	\begin{split}
	E(U_1^4)&=\pi_0mn+3\times\pi_1m(m-1)n+3\times\pi_9mn(n-1)+4\times\pi_1m(m-1)n+4\times\pi_9mn(n-1)\\&+6\times\pi_{12}mn(n-1)(n-2)+6\times\pi_2m(m-1)(m-2)n+12\times\pi_4m(m-1)n(n-1)\\&+12\times\pi_4m(m-1)n(n-1)+12\times\pi_4m(m-1)n(n-1)+6\times\pi_8m(m-1)n(n-1)\\&+4\times\pi_0^2m(m-1)n(n-1)+3\times\pi_0^2m(m-1)n(n-1)\\&+6\times\pi_0\pi_9m(m-1)n(n-1)(n-2)+6\times\pi_0\pi_1m(m-1)(m-2)n(n-1)\\&+12\times\pi_7m(m-1)n(n-1)(n-2)+12\times\pi_3m(m-1)(m-2)n(n-1)\\&+12\times\pi_0\pi_9m(m-1)n(n-1)(n-2)+12\times\pi_0\pi_1m(m-1)(m-2)n(n-1)\\&+12\times\pi_5m(m-1)(m-2)n(n-1)+12\times\pi_{10}m(m-1)n(n-1)(n-2)\\&+\pi_6mn(m-1)(m-2)(m-3)+\pi_{13}mn(n-1)(n-2)(n-3)\\&+4\times\pi_0\pi_{12}m(m-1)n(n-1)(n-2)(n-3)+4\times\pi_0\pi_2m(m-1)(m-2)(m-3)n(n-1)
	\\&+3\times\pi_{9}^2m(m-1)(m-2)(m-3)n(n-1)+3\times\pi_1^2m(m-1)n(n-1)(n-2)(n-3)\\&+6\times\pi_1\pi_{9}m(m-1)(m-2)n(n-1)(n-2)+6\times\pi_0^3m(m-1)(m-2)n(n-1)(n-2)\\&+24\times\pi_0\pi_4m(m-1)(m-2)n(n-1)(n-2)\\&+6\times\pi_0^2\pi_1m(m-1)(m-2)n(n-1)(n-2)(n-3)\\&+6\times\pi_9\pi_0^2m(m-1)(m-2)(m-3)n(n-1)(n-2)\\&+\pi_0^4m(m-1)(m-2)(m-3)n(n-1)(n-2)(n-3).
	\end{split}
	\end{align}
	Simplifying,
	\begin{align}
	\begin{split}
	E(U_1^4)&=\pi_0mn+7\times\pi_1m(m-1)n+7\times\pi_9mn(n-1)\\&+6\times\pi_{12}mn(n-1)(n-2)+6\times\pi_2m(m-1)(m-2)n+36\times\pi_4m(m-1)n(n-1)\\&+6\times\pi_8m(m-1)n(n-1)+7\times\pi_0^2m(m-1)n(n-1)\\&+6\times\pi_0\pi_9m(m-1)n(n-1)(n-2)+6\times\pi_0\pi_1m(m-1)(m-2)n(n-1)\\&+12\times\pi_7m(m-1)n(n-1)(n-2)+12\times\pi_3m(m-1)(m-2)n(n-1)\\&+12\times\pi_0\pi_9m(m-1)n(n-1)(n-2)+12\times\pi_0\pi_1m(m-1)(m-2)n(n-1)\\&+12\times\pi_5m(m-1)(m-2)n(n-1)+12\times\pi_{10}m(m-1)n(n-1)(n-2)\\&+\pi_6mn(m-1)(m-2)(m-3)+\pi_{13}mn(n-1)(n-2)(n-3)\\&+4\times\pi_0\pi_{12}m(m-1)n(n-1)(n-2)(n-3)+4\times\pi_0\pi_2m(m-1)(m-2)(m-3)n(n-1)
	\\&+3\times\pi_{9}^2m(m-1)(m-2)(m-3)n(n-1)+3\times\pi_1^2m(m-1)n(n-1)(n-2)(n-3)\\&+6\times\pi_1\pi_{9}m(m-1)(m-2)n(n-1)(n-2)+6\times\pi_0^3m(m-1)(m-2)n(n-1)(n-2)\\&+24\times\pi_0\pi_4m(m-1)(m-2)n(n-1)(n-2)\\&+6\times\pi_0^2\pi_1m(m-1)(m-2)n(n-1)(n-2)(n-3)\\&+6\times\pi_9\pi_0^2m(m-1)(m-2)(m-3)n(n-1)(n-2)\\&+\pi_0^4m(m-1)(m-2)(m-3)n(n-1)(n-2)(n-3).
	\end{split}
	\end{align}
	 As with $U_1$
	\begin{align}
	\begin{split}
	E(U_2^4)&=\pi_0MN+7\times\pi_1M(M-1)N+7\times\pi_9MN(N-1)\\&+6\times\pi_{12}MN(N-1)(N-2)+6\times\pi_2M(M-1)(M-2)N+36\times\pi_4M(M-1)N(N-1)\\&+6\times\pi_8M(M-1)N(N-1)+7\times\pi_0^2M(M-1)N(N-1)\\&+6\times\pi_0\pi_9M(M-1)N(N-1)(N-2)+6\times\pi_0\pi_1M(M-1)(M-2)N(N-1)\\&+12\times\pi_7M(M-1)N(N-1)(N-2)+12\times\pi_3M(M-1)(M-2)N(N-1)\\&+12\times\pi_0\pi_9M(M-1)N(N-1)(N-2)+12\times\pi_0\pi_1M(M-1)(M-2)N(N-1)\\&+12\times\pi_5M(M-1)(M-2)N(N-1)+12\times\pi_{10}M(M-1)N(N-1)(N-2)\\&+\pi_6MN(M-1)(M-2)(M-3)+\pi_{13}MN(N-1)(N-2)(N-3)\\&+4\times\pi_0\pi_{12}M(M-1)N(N-1)(N-2)(N-3)\\&+4\times\pi_0\pi_2M(M-1)(M-2)(M-3)N(N-1)
	\\&+3\times\pi_{9}^2M(M-1)(M-2)(M-3)N(N-1)\\&+3\times\pi_1^2M(M-1)N(N-1)(N-2)(N-3)\\&+6\times\pi_1\pi_{9}M(M-1)(M-2)N(N-1)(N-2)\\&+6\times\pi_0^3M(M-1)(M-2)N(N-1)(N-2)\\&+24\times\pi_0\pi_4M(M-1)(M-2)N(N-1)(N-2)\\&+6\times\pi_0^2\pi_1M(M-1)(M-2)N(N-1)(N-2)(N-3)\\&+6\times\pi_9\pi_0^2M(M-1)(M-2)(M-3)N(N-1)(N-2)\\&+\pi_0^4M(M-1)(M-2)(M-3)N(N-1)(N-2)(N-3).
	\end{split}
	\end{align}
	The conditional expectations are used to find mixed moments.
	\begin{align}
	\begin{split}
	&E(U_1U_2^3)=E(U_2^3E(U_1|U_2))=E(\frac{mn}{MN}U_2^4)=\frac{mn}{MN}E(U_2^4),\\
	&E(U_1U_2^3)=\pi_0mn+7\times\pi_1mn(M-1)+7\times\pi_9mn(N-1)\\&+6\times\pi_{12}mn(N-1)(N-2)+6\times\pi_2mn(M-1)(M-2)+36\times\pi_4mn(M-1)(N-1)\\&+6\times\pi_8mn(M-1)(N-1)+7\times\pi_0^2mn(M-1)(N-1)\\&+6\times\pi_0\pi_9mn(M-1)(N-1)(N-2)+6\times\pi_0\pi_1mn(M-1)(M-2)(N-1)\\&+12\times\pi_7mn(M-1)(N-1)(N-2)+12\times\pi_3mn(M-1)(M-2)(N-1)\\&+12\times\pi_0\pi_9mn(M-1)(N-1)(N-2)+12\times\pi_0\pi_1mn(M-1)(M-2)(N-1)\\&+12\times\pi_5mn(M-1)(M-2)(N-1)+12\times\pi_{10}mn(M-1)(N-1)(N-2)\\&+\pi_6mn(M-1)(M-2)(M-3)+\pi_{13}mn(N-1)(N-2)(N-3)\\&+4\times\pi_0\pi_{12}mn(M-1)(N-1)(N-2)(N-3)+4\times\pi_0\pi_2mn(M-1)(M-2)(M-3)(N-1)
	\\&+3\times\pi_{9}^2mn(M-1)(M-2)(M-3)(N-1)+3\times\pi_1^2mn(M-1)(N-1)(N-2)(N-3)\\&+6\times\pi_1\pi_{9}mn(M-1)(M-2)(N-1)(N-2)+6\times\pi_0^3mn(M-1)(M-2)(N-1)(N-2)\\&+24\times\pi_0\pi_4mn(M-1)(M-2)(N-1)(N-2)\\&+6\times\pi_0^2\pi_1mn(M-1)(M-2)(N-1)(N-2)(N-3)\\&+6\times\pi_9\pi_0^2mn(M-1)(M-2)(M-3)(N-1)(N-2)\\&+\pi_0^4mn(M-1)(M-2)(M-3)(N-1)(N-2)(N-3).
	\end{split}
	\end{align}

	$E(U_1^3U_2) $ and $ E(U_1^2U_2^2)$ are tough to find. We will spend some pages on it.

	\begin{align}
	\begin{split}
	E(U_1^2U_2^2)&=E(\sum\limits_{i=1}^m\sum\limits_{j=1}^n\sum\limits_{k=1}^m\sum\limits_{l=1}^nI_{ij}I_{kl}U_2^2),\\E(U_1^2U_2^2)&=E(\sum\limits_{i=1}^m\sum\limits_{j=1}^nI_{ij}U_2^2+\sum\limits_{i=1}^m\sum\limits_{j=1}^n\sum\limits_{k=1,k\neq i}^mI_{ij}I_{kj}U_2^2+\sum\limits_{i=1}^m\sum\limits_{j=1}^n\sum\limits_{l=1,l\neq j}^nI_{ij}I_{il}U_2^2\\&\quad+\sum\limits_{i=1}^m\sum\limits_{j=1}^n\sum\limits_{k=1,k\neq i}^m\sum\limits_{l=1,l\neq j}^nI_{ij}I_{kl}U_2^2).
	\end{split}
	\end{align}
	First, look at the fourth term\\
	$E(\sum\limits_{i=1}^m\sum\limits_{j=1}^n\sum\limits_{k=1,k\neq i}^m\sum\limits_{l=1,l\neq j}^nI_{ij}I_{kl}U_2^2)=m(m-1)n(n-1)E(I_{ij}I_{kl}U_2^2)$, for $ i\neq k$ and $ j\neq l$, $i,k\in\{1,2...M\}, j,l\in\{1,2...N\}$, because all of $ I_{ij}I_{kl}U_2^2$ are identical. And thus \\$E(\sum\limits_{i=1}^m\sum\limits_{j=1}^n\sum\limits_{k=1,k\neq i}^m\sum\limits_{l=1,l\neq j}^nI_{ij}I_{kl}U_2^2)=m(m-1)n(n-1)\frac{1}{M(M-1)N(N-1)}E(\sum\limits_{i=1}^M\sum\limits_{j=1}^N\sum\limits_{k=1,k\neq i}^M\sum\limits_{l=1,l\neq j}^NI_{ij}I_{kl}U_2^2)$,\\
	\begin{align}
	\begin{split}
	E(\sum\limits_{i=1}^m\sum\limits_{j=1}^n\sum\limits_{k=1,k\neq i}^m\sum\limits_{l=1,l\neq j}^nI_{ij}I_{kl}U_2^2)
	&=\frac{m(m-1)n(n-1)}{M(M-1)N(N-1)}E(\sum\limits_{i=1}^M\sum\limits_{j=1}^N\sum\limits_{k=1}^M\sum\limits_{l=1}^NI_{ij}I_{kl}U_2^2-\sum\limits_{i=1}^M\sum\limits_{j=1}^NI_{ij}U_2^2\\&-\sum\limits_{i=1}^M\sum\limits_{j=1}^N\sum\limits_{k=1,k\neq i}^MI_{ij}I_{kj}U_2^2-\sum\limits_{i=1}^M\sum\limits_{j=1}^N\sum\limits_{l=1,l\neq j}^NI_{ij}I_{il}U_2^2).
	\end{split}
	\end{align}
	Every $I_{ij}U_2^2$ is identical, where $i\in\{1,2...M\}, j\in\{1,2...N\}$. Every $I_{ij}I_{il}U_2^2$ is identical, where $i\in\{1,2...M\}, j\in\{1,2...N\},k\in\{1,2...M\}, k\neq i$. So is every $I_{ij}I_{kj}U_2^2$.
	\begin{align*}
		&E(\sum\limits_{i=1}^m\sum\limits_{j=1}^nI_{ij}U_2^2)=mnE(I_{ij}U_2^2)=\frac{mn}{MN}E(\sum\limits_{i=1}^M\sum\limits_{j=1}^NI_{ij}U_2^2),\\
		&E(\sum\limits_{i=1}^m\sum\limits_{j=1}^n\sum\limits_{k=1,k\neq i}^mI_{ij}I_{kj}U_2^2)=m(m-1)nE(I_{ij}I_{kj}U_2^2)=\frac{m(m-1)n}{M(M-1)N}E(\sum\limits_{i=1}^M\sum\limits_{j=1}^N\sum\limits_{k=1,k\neq i}^MI_{ij}I_{kj}U_2^2),\\
		&E(\sum\limits_{i=1}^m\sum\limits_{j=1}^n\sum\limits_{l=1,l\neq j}^nI_{ij}I_{il}U_2^2)=mn(n-1)E(I_{ij}I_{il}U_2^2)=\frac{mn(n-1)}{MN(N-1)}E(\sum\limits_{i=1}^M\sum\limits_{j=1}^N\sum\limits_{l=1,l\neq j}^NI_{ij}I_{il}U_2^2).
	\end{align*}
	Put them together, we get
	\begin{align}
	\begin{split}
    E(U_1^2U_2^2)=&(\frac{mn}{MN}-\frac{m(m-1)n(n-1)}{M(M-1)N(N-1)})E(\sum\limits_{i=1}^M\sum\limits_{j=1}^NI_{ij}U_2^2)\\&+(\frac{m(m-1)n}{M(M-1)N}-\frac{m(m-1)n(n-1)}{M(M-1)N(N-1)})E(\sum\limits_{i=1}^M\sum\limits_{j=1}^N\sum\limits_{k=1,k\neq i}^MI_{ij}I_{kj}U_2^2)\\&+(\frac{mn(n-1)}{MN(N-1)}-\frac{m(m-1)n(n-1)}{M(M-1)N(N-1)})E(\sum\limits_{i=1}^M\sum\limits_{j=1}^N\sum\limits_{l=1,l\neq j}^NI_{ij}I_{il}U_2^2)\\
    &+\frac{m(m-1)n(n-1)}{M(M-1)N(N-1)}E(\sum\limits_{i=1}^M\sum\limits_{j=1}^N\sum\limits_{k=1}^M\sum\limits_{l=1}^NI_{ij}I_{kl}U_2^2).
    \end{split}
	\end{align}
	 We know the following results, 
	\begin{align}
		E(\sum\limits_{i=1}^M\sum\limits_{j=1}^NI_{ij}U_2^2)=E(U_2^3),\\
		E(\sum\limits_{i=1}^M\sum\limits_{j=1}^N\sum\limits_{k=1}^M\sum\limits_{l=1}^NI_{ij}I_{kl}U_2^2)=E(U_2^4),
	\end{align}
	then the problem is to calculate $ E(\sum\limits_{i=1}^M\sum\limits_{j=1}^N\sum\limits_{k=1,k\neq i}^MI_{ij}I_{kj}U_2^2)$ and $E(\sum\limits_{i=1}^M\sum\limits_{j=1}^N\sum\limits_{l=1,l\neq j}^NI_{ij}I_{il}U_2^2) $.
	Let
	\begin{align*}
   H&=\sum\limits_{i=1}^M\sum\limits_{j=1}^N\sum\limits_{k=1,k\neq i}^MI_{ij}I_{kj}U_2^2\\
   &=\sum\limits_{i=1}^M\sum\limits_{j=1}^N\sum\limits_{k=1,k\neq i}^MI_{ij}I_{kj}U_{2,-i-k,-j}^2\\
   &+2\sum\limits_{i=1}^M\sum\limits_{j=1}^N\sum\limits_{k=1,k\neq i}^MI_{ij}I_{kj}U_{2,-i-k,-j}(\sum\limits_{a=1,a\neq i,a\neq k}I_{aj}+\sum\limits_{b=1,b\neq j}^N(I_{ib}+I_{kb})+I_{ij}+I_{kj})\\
   &+\sum\limits_{i=1}^M\sum\limits_{j=1}^N\sum\limits_{k=1,k\neq i}^MI_{ij}I_{kj}(\sum\limits_{a=1,a\neq i,a\neq k}^MI_{aj}+\sum\limits_{b=1,b\neq j}^N(I_{ib}+I_{kb})+I_{ij}+I_{kj})^2\\
  &=\sum\limits_{i=1}^M\sum\limits_{j=1}^N\sum\limits_{k=1,k\neq i}^MI_{ij}I_{kj}U_{2,-i-k,-j}^2\\
   &+2\sum\limits_{i=1}^M\sum\limits_{j=1}^N\sum\limits_{k=1,k\neq i}^MI_{ij}I_{kj}U_{2,-i-k,-j}(\sum\limits_{a=1,a\neq i,a\neq k}I_{aj}+\sum\limits_{b=1,b\neq j}^N(I_{ib}+I_{kb})+I_{ij}+I_{kj})\\
   &+\sum\limits_{i=1}^M\sum\limits_{j=1}^N\sum\limits_{k=1,k\neq i}^MI_{ij}I_{kj}(\sum\limits_{a=1,a\neq i,a\neq k}^MI_{aj}+\sum\limits_{b=1,b\neq j}^N(I_{ib}+I_{kb})+I_{ij}+I_{kj})^2
   \\
   &=\sum\limits_{i=1}^M\sum\limits_{j=1}^N\sum\limits_{k=1,k\neq i}^MI_{ij}I_{kj}U_{2,-i-k,-j}^2\\
   &+2\sum\limits_{i=1}^M\sum\limits_{j=1}^N\sum\limits_{k=1,k\neq i}^MI_{ij}I_{kj}U_{2,-i-k,-j}\sum\limits_{a=1,a\neq i,a\neq k}I_{aj}\\&+2\sum\limits_{i=1}^M\sum\limits_{j=1}^N\sum\limits_{k=1,k\neq i}^MI_{ij}I_{kj}U_{2,-i-k,-j}(\sum\limits_{b=1,b\neq j}^N(I_{ib}+I_{kb})+2)\\
   &+\sum\limits_{i=1}^M\sum\limits_{j=1}^N\sum\limits_{k=1,k\neq i}^MI_{ij}I_{kj}(\sum\limits_{a=1,a\neq i,a\neq k}^MI_{aj}+\sum\limits_{b=1,b\neq j}^N(I_{ib}+I_{kb})+2)^2\\
   &=\sum\limits_{i=1}^M\sum\limits_{j=1}^N\sum\limits_{k=1,k\neq i}^MI_{ij}I_{kj}U_{2,-i-k,-j}^2\\
   &+2\sum\limits_{i=1}^M\sum\limits_{j=1}^N\sum\limits_{k=1,k\neq i}^M\sum\limits_{a=1,a\neq i,a\neq k}^MI_{aj}I_{ij}I_{kj}(U_{2,-i-k-a,-j}+\sum\limits_{b=1,b\neq j}^NI_{ab})\\&+2\sum\limits_{i=1}^M\sum\limits_{j=1}^N\sum\limits_{k=1,k\neq i}^M\sum\limits_{b=1,b\neq j}^NI_{ij}I_{kj}U_{2,-i-k,-j}(I_{ib}+I_{kb})\\
   &+4\sum\limits_{i=1}^M\sum\limits_{j=1}^N\sum\limits_{k=1,k\neq i}^MI_{ij}I_{kj}U_{2,-i-k,-j}\\
   &+\sum\limits_{i=1}^M\sum\limits_{j=1}^N\sum\limits_{k=1,k\neq i}^MI_{ij}I_{kj}(\sum\limits_{a=1,a\neq i,a\neq k}^MI_{aj}+\sum\limits_{b=1,b\neq j}^N(I_{ib}+I_{kb})+2)^2,
	\end{align*}
	\begin{align*}
	H=&\sum\limits_{i=1}^M\sum\limits_{j=1}^N\sum\limits_{k=1,k\neq i}^MI_{ij}I_{kj}U_{2,-i-k,-j}^2\\
	&+2\sum\limits_{i=1}^M\sum\limits_{j=1}^N\sum\limits_{k=1,k\neq i}^M\sum\limits_{a=1,a\neq i,a\neq k}^MI_{aj}I_{ij}I_{kj}U_{2,-i-k-a,-j}\\&+2\sum\limits_{i=1}^M\sum\limits_{j=1}^N\sum\limits_{k=1,k\neq i}^M\sum\limits_{a=1,a\neq i,a\neq k}^M\sum\limits_{b=1,b\neq j}^NI_{aj}I_{ij}I_{kj}I_{ab}\\&+2\sum\limits_{i=1}^M\sum\limits_{j=1}^N\sum\limits_{k=1,k\neq i}^M\sum\limits_{b=1,b\neq j}^NI_{ij}I_{kj}U_{2,-i-k,-j}(I_{ib}+I_{kb})\\
	&+4\sum\limits_{i=1}^M\sum\limits_{j=1}^N\sum\limits_{k=1,k\neq i}^MI_{ij}I_{kj}U_{2,-i-k,-j}\\
	&+\sum\limits_{i=1}^M\sum\limits_{j=1}^N\sum\limits_{k=1,k\neq i}^M\sum\limits_{a=1,a\neq i,a\neq k}^M\sum\limits_{c=1,c\neq i,c\neq k}^MI_{ij}I_{kj}I_{aj}I_{cj}\\
	&+\sum\limits_{i=1}^M\sum\limits_{j=1}^N\sum\limits_{k=1,k\neq i}^M\sum\limits_{b=1,b\neq j}^N\sum\limits_{d=1,d\neq j}^NI_{ij}I_{kj}(I_{ib}+I_{kb})(I_{id}+I_{kd})\\
	&+4\sum\limits_{i=1}^M\sum\limits_{j=1}^N\sum\limits_{k=1,k\neq i}^MI_{ij}I_{kj}\\
	&+4\sum\limits_{i=1}^M\sum\limits_{j=1}^N\sum\limits_{k=1,k\neq i}^M\sum\limits_{a=1,a\neq i,a\neq k}^MI_{ij}I_{kj}I_{aj}\\
	&+4\sum\limits_{i=1}^M\sum\limits_{j=1}^N\sum\limits_{k=1,k\neq i}^M\sum\limits_{b=1,b\neq j}^NI_{ij}I_{kj}(I_{ib}+I_{kb})\\
	&+2\sum\limits_{i=1}^M\sum\limits_{j=1}^N\sum\limits_{k=1,k\neq i}^M\sum\limits_{a=1,a\neq i,a\neq k}^M\sum\limits_{b=1,b\neq j}^NI_{aj}I_{ij}I_{kj}(I_{ib}+I_{kb}),
	\end{align*}
		\begin{align*}
	H=&\sum\limits_{i=1}^M\sum\limits_{j=1}^N\sum\limits_{k=1,k\neq i}^MI_{ij}I_{kj}U_{2,-i-k,-j}^2\\
	&+2\sum\limits_{i=1}^M\sum\limits_{j=1}^N\sum\limits_{k=1,k\neq i}^M\sum\limits_{a=1,a\neq i,a\neq k}^MI_{aj}I_{ij}I_{kj}U_{2,-i-k-a,-j}\\&+2\sum\limits_{i=1}^M\sum\limits_{j=1}^N\sum\limits_{k=1,k\neq i}^M\sum\limits_{a=1,a\neq i,a\neq k}^M\sum\limits_{b=1,b\neq j}^NI_{aj}I_{ij}I_{kj}I_{ab}\\&+2\sum\limits_{i=1}^M\sum\limits_{j=1}^N\sum\limits_{k=1,k\neq i}^M\sum\limits_{b=1,b\neq j}^NI_{ij}I_{kj}U_{2,-i-k,-j-b}(I_{ib}+I_{kb})\\
	&+2\sum\limits_{i=1}^M\sum\limits_{j=1}^N\sum\limits_{k=1,k\neq i}^M\sum\limits_{b=1,b\neq j}^N\sum\limits_{a=1,a\neq i,a\neq k}^MI_{ij}I_{kj}I_{ab}(I_{ib}+I_{kb})\\
	&+4\sum\limits_{i=1}^M\sum\limits_{j=1}^N\sum\limits_{k=1,k\neq i}^MI_{ij}I_{kj}U_{2,-i-k,-j}\\
	&+\sum\limits_{i=1}^M\sum\limits_{j=1}^N\sum\limits_{k=1,k\neq i}^M\sum\limits_{a=1,a\neq i,a\neq k}^M\sum\limits_{c=1,c\neq a,c\neq i,c\neq k}^MI_{ij}I_{kj}I_{aj}I_{cj}\\
	&+\sum\limits_{i=1}^M\sum\limits_{j=1}^N\sum\limits_{k=1,k\neq i}^M\sum\limits_{a=1,a\neq i,a\neq k}^MI_{ij}I_{kj}I_{aj}\\
	&+2\sum\limits_{i=1}^M\sum\limits_{j=1}^N\sum\limits_{k=1,k\neq i}^M\sum\limits_{b=1,b\neq j}^N\sum\limits_{d=1,d\neq j,d\neq b}^NI_{ij}I_{kj}(I_{ib}I_{id}+I_{id}I_{kb})\\
	&+2\sum\limits_{i=1}^M\sum\limits_{j=1}^N\sum\limits_{k=1,k\neq i}^M\sum\limits_{b=1,b\neq j}^NI_{ij}I_{kj}(I_{ib}+I_{ib}I_{kb})\\
	&+4\sum\limits_{i=1}^M\sum\limits_{j=1}^N\sum\limits_{k=1,k\neq i}^MI_{ij}I_{kj}\\
	&+4\sum\limits_{i=1}^M\sum\limits_{j=1}^N\sum\limits_{k=1,k\neq i}^M\sum\limits_{a=1,a\neq i,a\neq k}^MI_{ij}I_{kj}I_{aj}\\
	&+4\sum\limits_{i=1}^M\sum\limits_{j=1}^N\sum\limits_{k=1,k\neq i}^M\sum\limits_{b=1,b\neq j}^NI_{ij}I_{kj}(I_{ib}+I_{kb})\\
	&+2\sum\limits_{i=1}^M\sum\limits_{j=1}^N\sum\limits_{k=1,k\neq i}^M\sum\limits_{a=1,a\neq i,a\neq k}^M\sum\limits_{b=1,b\neq j}^NI_{aj}I_{ij}I_{kj}(I_{ib}+I_{kb}),
	\end{align*}
	\begin{align*}
	E(H)=&\sum\limits_{i=1}^M\sum\limits_{j=1}^N\sum\limits_{k=1,k\neq i}^M\pi_1E(U_{2,-i-k,-j}^2)\\
	&+2\sum\limits_{i=1}^M\sum\limits_{j=1}^N\sum\limits_{k=1,k\neq i}^M\sum\limits_{a=1,a\neq i,a\neq k}^M\pi_2E(U_{2,-i-k-a,-j})\\
	&+2\sum\limits_{i=1}^M\sum\limits_{j=1}^N\sum\limits_{k=1,k\neq i}^M\sum\limits_{a=1,a\neq i,a\neq k}^M\sum\limits_{b=1,b\neq j}^N\pi_3\\
	&+2\sum\limits_{i=1}^M\sum\limits_{j=1}^N\sum\limits_{k=1,k\neq i}^M\sum\limits_{b=1,b\neq j}^N2\pi_4E(U_{2,-i-k,-j-b})\\
	&+2\sum\limits_{i=1}^M\sum\limits_{j=1}^N\sum\limits_{k=1,k\neq i}^M\sum\limits_{b=1,b\neq j}^N\sum\limits_{a=1,a\neq i,a\neq k}^M2\pi_5\\
	&+4\sum\limits_{i=1}^M\sum\limits_{j=1}^N\sum\limits_{k=1,k\neq i}^M\pi_1E(U_2,-i-k,-j)\\
	&+\sum\limits_{i=1}^M\sum\limits_{j=1}^N\sum\limits_{k=1,k\neq i}^M\sum\limits_{a=1,a\neq i,a\neq k}\pi_2\\
	&+\sum\limits_{i=1}^M\sum\limits_{j=1}^N\sum\limits_{k=1,k\neq i}^M\sum\limits_{a=1,a\neq i,a\neq k}\sum\limits_{c=1,c\neq a,c\neq i,c\neq k}\pi_6\\
	&+2\sum\limits_{i=1}^M\sum\limits_{j=1}^N\sum\limits_{k=1,k\neq i}^M\sum\limits_{b=1,b\neq j}(\pi_4+\pi_8)\\
	&+2\sum\limits_{i=1}^M\sum\limits_{j=1}^N\sum\limits_{k=1,k\neq i}^M\sum\limits_{b=1,b\neq j}\sum\limits_{d=1,d\neq j,d\neq b}(\pi_7+\pi_{10})\\
	&+4\sum\limits_{i=1}^M\sum\limits_{j=1}^N\sum\limits_{k=1,k\neq i}^M\pi_1\\
	&+4\sum\limits_{i=1}^M\sum\limits_{j=1}^N\sum\limits_{k=1,k\neq i}^M\sum\limits_{a=1,a\neq i,a\neq k}^M\pi_2\\
	&+4\sum\limits_{i=1}^M\sum\limits_{j=1}^N\sum\limits_{k=1,k\neq i}^M\sum\limits_{b=1,b\neq j}^N2\pi_4\\
	&+2\sum\limits_{i=1}^M\sum\limits_{j=1}^N\sum\limits_{k=1,k\neq i}^M\sum\limits_{a=1,a\neq i,a\neq k}^M\sum\limits_{b=1,b\neq j}^N2\pi_3.
	\end{align*}
    \begin{align}
    \begin{split}
    E(H)=&M(M-1)N\{\pi_1(M-2)(N-1)[\pi_0+(M-3)\pi_1+(N-2)\pi_9+(M-3)(N-2)\pi_0^2]\\
    &+2(M-2)(M-3)(N-1)\pi_2\pi_0+2(M-2)(N-1)\pi_3\\
    &+4(N-1)(M-2)(N-2)\pi_4\pi_0+4(N-1)(M-2)\pi_5\\
    &+4(M-2)(N-1)\pi_1\pi_0+(M-2)\pi_2\\
    &+(M-2)(M-3)\pi_6+2(N-1)(\pi_4+\pi_8)\\
    &+2(N-1)(N-2)(\pi_7+\pi_{10})\\
    &+4\pi_1+4(M-2)\pi_2\\
    &+8(N-1)\pi_4+4(M-2)(N-1)\pi_3\}.
    \end{split}
    \end{align}
    By the same reasoning,
     \begin{align}
    \begin{split}
    E(K)=&N(N-1)M\{\pi_1(N-2)(M-1)[\pi_0+(N-3)\pi_1+(M-2)\pi_9+(N-3)(M-2)\pi_0^2]\\
    &+2(N-2)(N-3)(M-1)\pi_2\pi_0+2(N-2)(M-1)\pi_3\\
    &+4(M-1)(N-2)(M-2)\pi_4\pi_0+4(M-1)(N-2)\pi_5\\
    &+4(N-2)(M-1)\pi_1\pi_0+(N-2)\pi_2\\
    &+(N-2)(N-3)\pi_6+2(M-1)(\pi_4+\pi_8)\\
    &+2(M-1)(M-2)(\pi_7+\pi_{10})\\
    &+4\pi_1+4(N-2)\pi_2\\
    &+8(M-1)\pi_4+4(N-2)(M-1)\pi_3\}.
    \end{split}
    \end{align}
    By (38), (39), (40), (41) and (42), the result is done. The same method is used to find $E(U_1^3U_2)$. Expand it and then calculate each part.
    \begin{align}
    \begin{split}
    E(U_1^3U_2)=&E(\sum\limits_{i=1}^m\sum\limits_{j=1}^n\sum\limits_{k=1}^m\sum\limits_{j=1}^n\sum\limits_{s=1}^m\sum\limits_{t=1}^nI_{ij}I_{kl}I_{st}U_2)
    \\&=E(\sum\limits_{i=1}^m\sum\limits_{j=1}^n\sum\limits_{k=1,k \neq i}^m\sum\limits_{l=1, l\neq j}^n\sum\limits_{s=1,s\neq i, s\neq k}^m\sum\limits_{t=1,t\neq j,t\neq l}^nI_{ij}I_{kl}I_{st}U_2)
    \\&+3E(\sum\limits_{i=1}^m\sum\limits_{j=1}^n\sum\limits_{k=1,k \neq i}^m\sum\limits_{l=1, l\neq j}^n\sum\limits_{t=1,t\neq j,t\neq l}^nI_{ij}I_{kl}I_{it}U_2)\\&+3E(\sum\limits_{i=1}^m\sum\limits_{j=1}^n\sum\limits_{k=1,k \neq i}^m\sum\limits_{l=1, l\neq j}^n\sum\limits_{s=1,s\neq i, s\neq k}^mI_{ij}I_{kl}I_{sl}U_2)\\&+E(\sum\limits_{i=1}^m\sum\limits_{j=1}^n\sum\limits_{l=1, l\neq j}^n\sum\limits_{t=1,t\neq j,t\neq l}^nI_{ij}I_{il}I_{it}U_2)+E(\sum\limits_{i=1}^m\sum\limits_{j=1}^n\sum\limits_{k=1,k \neq i}^m\sum\limits_{s=1,s\neq i, s\neq k}^mI_{ij}I_{kj}I_{sj}U_2)\\&+6E(\sum\limits_{i=1}^m\sum\limits_{j=1}^n\sum\limits_{k=1,k \neq i}^m\sum\limits_{j=1, j\neq l}^nI_{ij}I_{il}I_{kj}U_2)+3E(\sum\limits_{i=1}^m\sum\limits_{j=1}^n\sum\limits_{k=1,k \neq i}^m\sum\limits_{l=1, l\neq j}^nI_{ij}I_{kl}U_2)\\
    &+3E(\sum\limits_{i=1}^m\sum\limits_{j=1}^n\sum\limits_{k=1,k \neq i}^mI_{ij}I_{kj}U_2)+3E(\sum\limits_{i=1}^m\sum\limits_{j=1}^n\sum\limits_{l=1, l\neq j}^nI_{ij}I_{il}U_2)
    +E(\sum\limits_{i=1}^m\sum\limits_{j=1}^nI_{ij}U_2).
    \end{split}
    \end{align}
    
    \begin{align*}
    &E(U_1^3U_2)\\=&\frac{m(m-1)(m-2)n(n-1)(n-2)}{M(M-1)(M-2)N(N-1)(N-2)}E(\sum\limits_{i=1}^M\sum\limits_{j=1}^N\sum\limits_{k=1,k \neq i}^M\sum\limits_{l=1, l\neq j}^N\sum\limits_{s=1,s\neq i, s\neq k}^M\sum\limits_{t=1,t\neq j,t\neq l}^NI_{ij}I_{kl}I_{st}U_2)\\&+3\frac{m(m-1)n(n-1)(n-2)}{M(M-1)N(N-1)(N-2)}E(\sum\limits_{i=1}^M\sum\limits_{j=1}^N\sum\limits_{k=1,k \neq i}^M\sum\limits_{l=1, l\neq j}^N\sum\limits_{t=1,t\neq j,t\neq l}^NI_{ij}I_{kl}I_{it}U_2)\\
    &+3\frac{m(m-1)(m-2)n(n-1)}{M(M-1)(M-2)N(N-1)}E(\sum\limits_{i=1}^M\sum\limits_{j=1}^N\sum\limits_{k=1,k \neq i}^M\sum\limits_{l=1, l\neq j}^N\sum\limits_{s=1,s\neq i, s\neq k}^MI_{ij}I_{kl}I_{sl}U_2)
    \\&+\frac{mn(n-1)(n-2)}{MN(N-1)(N-2)}E(\sum\limits_{i=1}^M\sum\limits_{j=1}^N\sum\limits_{l=1, l\neq j}^N\sum\limits_{t=1,t\neq j,t\neq l}^NI_{ij}I_{il}I_{it}U_2)\\
    &+\frac{m(m-1)(m-2)n}{M(M-1)(M-2)N}E(\sum\limits_{i=1}^M\sum\limits_{j=1}^N\sum\limits_{k=1,k \neq i}^M\sum\limits_{s=1,s\neq i, s\neq k}^MI_{ij}I_{kj}I_{sj}U_2)\\
    &+6\frac{m(m-1)n(n-1)}{M(M-1)N(N-1)}E(\sum\limits_{i=1}^M\sum\limits_{j=1}^N\sum\limits_{k=1,k \neq i}^M\sum\limits_{l=1, l\neq j}^NI_{ij}I_{il}I_{kj}U_2)\\
    &+3\frac{m(m-1)n(n-1)}{M(M-1)N(N-1)}E(\sum\limits_{i=1}^m\sum\limits_{j=1}^n\sum\limits_{k=1,k \neq i}^m\sum\limits_{l=1, l\neq j}^nI_{ij}I_{kl}U_2)\\
    &+3\frac{m(m-1)n}{M(M-1)N}E(\sum\limits_{i=1}^M\sum\limits_{j=1}^N\sum\limits_{k=1,k \neq i}^MI_{ij}I_{kj}U_2)+3\frac{mn(n-1)}{MN(N-1)}E(\sum\limits_{i=1}^M\sum\limits_{j=1}^N\sum\limits_{l=1, l\neq j}^NI_{ij}I_{il}U_2)\\&+\frac{mn}{MN}E(\sum\limits_{i=1}^M\sum\limits_{j=1}^NI_{ij}U_2)\\
    =&\frac{m(m-1)(m-2)n(n-1)(n-2)}{M(M-1)(M-2)N(N-1)(N-2)}M(M-1)(M-2)N(N-1)(N-2)[\pi_0^4(M-3)(N-3)\\&+3(N-3)\pi_0^2\pi_9+3(M-3)\pi_1\pi_0^2+3\pi_0^3+6\pi_0\pi_4)]\\
    &+3\frac{m(m-1)n(n-1)(n-2)}{M(M-1)N(N-1)(N-2)}M(M-1)N(N-1)(N-2)
    [\pi_0^2\pi_9(M-2)(N-3)+\pi_0\pi_{12}(N-3)\\&+\pi_9^2(N-3)+\pi_9\pi_1(M-2)+2\pi_4\pi_0(M-2)+
    \pi_9\pi_0+\pi_7+\pi_9\pi_0+\pi_{10}+\pi_0\pi_9+\pi_{10}]\\
    &+3\frac{m(m-1)n(n-1)(n-2)}{M(M-1)N(N-1)(N-2)}(M(M-1)N(N-1)(N-2))
    [\pi_0^2\pi_1(M-2)(N-3)\\&+\pi_0\pi_2(N-3)+\pi_1^2(N-3)+\pi_9\pi_1(M-2)+2\pi_4\pi_0
    \pi_1\pi_0+\pi_3+\pi_1\pi_0+\pi_5+\pi_0\pi_1+\pi_5]\\
    &+\frac{mn(n-1)(n-2)}{MN(N-1)(N-2)}
    MN(N-1)(N-2)[(M-1)(N-3)\pi_0\pi_{12}+(N-3)\pi_{13}+3(M-1)\pi_7+3\pi_{12}]\\
    &+\frac{m(m-1)(m-2)n}{M(M-1)(M-2)N}
    MN(M-1)(M-2)[(N-1)(M-3)\pi_0\pi_2+(M-3)\pi_6+(N-1)\pi_33+3\pi_2]
    \\&+
    6\frac{m(m-1)n(n-1)}{M(M-1)N(N-1)}M(M-1)N(N-1)
    [(M-2)(N-2)\pi_0\pi_4+(N-2)\pi_7+(N-2)\pi_10\\&+(M-2)\pi_3+(M-2)\pi_5+3\pi_4+\pi_8]
    \end{align*}
    \begin{align*}
    \\&+
    \frac{3m(m-1)n(n-1)}{M(M-1)N(N-1)}M(M-1)N(N-1)
    [(M-2)(N-2)\pi_0^3+2(N-2)\pi_0\pi_9+(M-2)\pi_0\pi_{12}+\\&\quad2\pi_0^2+2\pi_4]
    \\&+
    \frac{3m(m-1)n}{M(M-1)N}M(M-1)N
    [(M-2)(N-1)\pi_0\pi_1+(N-1)\pi_42+(M-2)\pi_2+2\pi_1]
    \\&+
    \frac{3mn(n-1)}{MN(N-1)}
    M(N-1)N[(N-2)(M-1)\pi_0\pi_9+(M-1)\pi_42+(N-2)\pi_12+2\pi_9]
    \\&+\frac{mn}{MN}E(U_2^2).
    \end{align*}
    Simplifying,
    \begin{align}
    \begin{split}
   E(U_1^3U_2)=&m(m-1)(m-2)n(n-1)(n-2)[\pi_0^4(M-3)(N-3)+3(N-3)\pi_0^2\pi_9+3(M-3)\pi_1\pi_0^2+\\&3\pi_0^3+6\pi_0\pi_4)]\\
    &+3m(m-1)n(n-1)(n-2)[\pi_0^2\pi_9(M-2)(N-3)+\pi_0\pi_{12}(N-3)+\pi_9^2(N-3)+\\&\pi_9\pi_1(M-2)+2\pi_4\pi_0(M-2)+
    \pi_9\pi_0+\pi_7+\pi_9\pi_0+\pi_{10}+\pi_0\pi_9+\pi_{10}]\\
    &+3m(m-1)n(n-1)(n-2)
    [\pi_0^2\pi_1(M-2)(N-3)+\pi_0\pi_2(N-3)+\pi_1^2(N-3)+\\&\pi_9\pi_1(M-2)+2\pi_4\pi_0
    \pi_1\pi_0+\pi_3+\pi_1\pi_0+\pi_5+\pi_0\pi_1+\pi_5]\\
    &+mn(n-1)(n-2)[(M-1)(N-3)\pi_0\pi_{12}+(N-3)\pi_{13}+3(M-1)\pi_7+3\pi_{12}]\\
    &+m(m-1)(m-2)n[(N-1)(M-3)\pi_0\pi_2+(M-3)\pi_6+(N-1)\pi_33+3\pi_2]
    \\&+
    6m(m-1)n(n-1) [(M-2)(N-2)\pi_0\pi_4+(N-2)\pi_7+(N-2)\pi_10+\\&(M-2)\pi_3+(M-2)\pi_5+3\pi_4+\pi_8]
    \\&+
    3m(m-1)n(n-1)
    [(M-2)(N-2)\pi_0^3+2(N-2)\pi_0\pi_9+(M-2)\pi_0\pi_{12}+\\&\quad2\pi_0^2+2\pi_4]
    \\&+
    3m(m-1)n
    [(M-2)(N-1)\pi_0\pi_1+(N-1)\pi_42+(M-2)\pi_2+2\pi_1]
    \\&+
    3mn(n-1)[(N-2)(M-1)\pi_0\pi_9+(M-1)\pi_42+(N-2)\pi_12+2\pi_9]
    \\&+\frac{mn}{MN}E(U_2^2).
    \end{split}
    \end{align}
    
	\par Under the null hypothesis,
	\begin{align*}
	E(U_1^4)&=\frac{1}{2}mn+3\times\frac{1}{3}m(m-1)n+3\times\frac{1}{3}mn(n-1)+4\times\frac{1}{3}m(m-1)n+4\times\frac{1}{3}mn(n-1)\\&+6\times\frac{1}{4}mn(n-1)(n-2)+6\times\frac{1}{4}m(m-1)(m-2)n+12\times\frac{5}{24}m(m-1)n(n-1)\\&+12\times\frac{2}{9}m(m-1)n(n-1)+12\times\frac{5}{24}+6\times\frac{1}{6}m(m-1)n(n-1)+4\times\frac{1}{4}m(m-1)n(n-1)\\&+3\times\frac{1}{4}m(m-1)n(n-1)+6\times\frac{1}{6}m(m-1)n(n-1)(n-2)+6\times\frac{1}{6}m(m-1)(m-2)n(n-1)\\&+12\times\frac{3}{20}m(m-1)n(n-1)(n-2)+12\times\frac{3}{20}m(m-1)(m-2)n(n-1)\\&+12\times\frac{1}{6}m(m-1)n(n-1)(n-2)+12\times\frac{1}{6}m(m-1)(m-2)n(n-1)\\&+12\times\frac{2}{15}m(m-1)(m-2)n(n-1)+12\times\frac{2}{15}m(m-1)n(n-1)(n-2)\\&+4\times\frac{1}{8}m(m-1)n(n-1)(n-2)(n-3)+4\times\frac{1}{8}m(m-1)(m-2)(m-3)n(n-1)
	\\&+3\times\frac{1}{9}m(m-1)(m-2)(m-3)n(n-1)+3\times\frac{1}{9}m(m-1)n(n-1)(n-2)(n-3)\\&+6\times\frac{1}{9}m(m-1)(m-2)n(n-1)(n-2)+6\times\frac{1}{8}m(m-1)(m-2)n(n-1)(n-2)\\&+24\times\frac{5}{48}m(m-1)(m-2)n(n-1)(n-2)+6\times\frac{1}{12}m(m-1)(m-2)n(n-1)(n-2)(n-3)\\&+6\times\frac{1}{12}m(m-1)(m-2)(m-3)n(n-1)(n-2)\\&+\frac{1}{16}m(m-1)(m-2)(m-3)n(n-1)(n-2)(n-3).
	\end{align*}
	\begin{align*}
	E(U_1^4)=& m^4n^4/16 + m^4n^3/8 + m^4n^2/48 - m^4n/120 + m^3n^4/8 +
	m^3n^3/6 + m^3n^2/40\\& - m^3n/60 + m^2n^4/48 + m^2n^3/40 - 
	m^2n^2/240 - m^2n/120 - mn^4/120 - mn^3/60 \\&- mn^2/120.
	\end{align*}
	\begin{align*}
	E(U_2^4)=& M^4N^4/16 + M^4N^3/8 + M^4N^2/48 - M^4N/120 + M^3N^4/8 +
	M^3N^3/6 + M^3N^2/40\\& - M^3N/60 + M^2N^4/48 + M^2N^3/40 - 
	M^2N^2/240 - M^2N/120 - MN^4/120 - MN^3/60 \\&- MN^2/120.
	\end{align*}
	\begin{align*}
	&E(U_1U_2^3)=E(U_2^3E(U_1|U_2))=E(\frac{mn}{MN}U_2^4),\\
	&E(U_1U_2^3)=mnM^3N^3/16+mnM^3N^2/8+mnM^3N/48 - mnM^3/120 + mnM^2N^3/8\\&\qquad \qquad \quad+
	mnM^2N^2/6 +mn M^2N/40-mn M^2/60 + mnMN^3/48 + mnMN^2/40 - 
	mnMN/240 \\&\qquad \qquad\quad- mnM/120 -mn N^3/120 - mnN^2/60 - mnN/120.
	\end{align*}
	\begin{align*}
	&E(U_1^3U_2)=E(U_1^3E(U_2|U_1)),\\&E(U_1^3U_2)=E[U_1^3(\frac{M+N+1}{m+n+1}U_1+\frac{1}{2}[\frac{(M-m)n(n+1)+(N-n)m(m+1)}{m+n+1}\\&\qquad \quad \qquad+(M-m)(N-n)])]\\&\qquad \qquad=\frac{M+N+1}{m+n+1}E(U_1^4)+\frac{1}{2}\frac{(M-m)n(n+1)+(N-n)m(m+1)}{m+n+1}E(U_1^3)\\&\qquad\qquad\quad+\frac{1}{2}(M-m)(N-n)E(U_1^3).\\
	\end{align*}	
	\begin{align*}
	E(H)=E(\sum\limits_{i=1}^M\sum\limits_{j=1}^N\sum\limits_{k=1,k\neq i}^MI_{ij}I_{kj}U_2^2)&=MN(M - 1)\{5M/4 + 29N/12 + 3\times(2M - 4)(N - 1)/20 \\&+ 2\times(4N - 4)(M - 2)/15 + 19\times(4M - 8)(N - 1)/60 +\\& 17\times(2N - 2)(N - 2)/60 + (M - 2)(M - 3)/5 + \\&(M/3 - 2/3)(N - 1)(M/3 + N/3 + (M - 3)(N - 2)/4 - 7/6) \\&+ (2M - 4)(M - 3)(N - 1)/8 + \\&5\times(4N - 4)(M - 2)(N - 2)/48 - 43/12\}.
	\end{align*}
	\begin{align*}
	E(G)=E(\sum\limits_{j=1}^N\sum\limits_{i=1}^M\sum\limits_{l=1,l\neq j}^NI_{ij}I_{il}U_2^2)&=MN(N - 1)\{5N/4 + 29M/12 + 3\times(2N - 4)(M - 1)/20 \\&+ 2\times(4M - 4)(N - 2)/15 + 19\times(4N - 8)(M - 1)/60 +\\& 17\times(2M - 2)(M - 2)/60 + (N - 2)(N - 3)/5\\& + (N/3 - 2/3)(M - 1)(N/3 + M/3 + (N - 3)(M - 2)/4 - 7/6) \\&+ (2N - 4)(N - 3)(M - 1)/8 + \\&5\times(4M - 4)(N - 2)(M - 2)/48 - 43/12\}.\\
	\end{align*}
	\begin{align*}
	\begin{split}
	 E(U_1^2U_2^2)=&(\frac{mn}{MN}-\frac{m(m-1)n(n-1)}{M(M-1)N(N-1)})E(U_2^3)\\&+(\frac{m(m-1n)}{M(M-1)N}-\frac{m(m-1)n(n-1)}{M(M-1)N(N-1)})E(H)\\&+(\frac{mn(n-1)}{MN(N-1)}-\frac{m(m-1)n(n-1)}{M(M-1)N(N-1)})E(G)\\
	&+\frac{m(m-1)n(n-1)}{M(M-1)N(N-1)}E(U_2^4).
	\end{split}
	\end{align*}
	\section{Cumulants}
	In this section, we will find the relation between the cumulants and moments. So we can use the moments to calculate the cumulants which we use to substitute in Cornish Fisher expansion to get the critical values.
	As the first step, we express derivatives of the cumulant generating function in terms of derivatives of the moment generating function.\\
	The following are moment generating function and cumulant generating function of $(U_1,U_2)$ \\
	$M(t_1,t_2)=E(e^{U_1t_1+U_2t_2}) \qquad K(t_1,t_2)=logE(e^{t_1U_1+t_2U_2}) \\	$
	For short notations, we denote$ \quad M=E(e^{U_1t_1+U_2t_2}), \qquad S=e^{t_1U_1+t_2U_2}. \\ $
	\subsection{1st derivatives}
	$\frac{\partial K(t_1,t_2)}{\partial t_1}=\frac{\partial logM}{\partial t_1}=M^{-1}\frac{\partial M}{\partial t_1}=M^{-1}E(U_1S), \\ $
	$\frac{\partial K(t_1,t_2)}{\partial t_2}=\frac{\partial logM}{\partial t_2}=M^{-1}\frac{\partial M}{\partial t_2}=M^{-1}E(U_2S). \\  $
	\subsection{2nd derivatives}
	$\frac{\partial^2 K(t_1,t_2)}{\partial t_{1}^2}=M^{-1}\frac{\partial E(U_1S)}{\partial t_1}+E(U_1S)\frac{\partial M^{-1}}{\partial t_1}=M^{-1}E(U_1^2S)-M^{-2}E(U_1S)E(U_1S), \\ $
	$\frac{\partial^2 K(t_1,t_2)}{\partial t_{2}^2}=M^{-1}\frac{\partial E(U_2S)}{\partial t_2}+E(U_2S)\frac{\partial M^{-1}}{\partial t_2}=M^{-1}E(U_2^2S)-M^{-2}E(U_2S)E(U_2S),\\   $
	$\frac{\partial^2 K(t_1,t_2)}{\partial t_1t_{2}}=M^{-1}\frac{\partial E(U_1S)}{\partial t_2}+E(U_1S)\frac{\partial M^{-1}}{\partial t_2}=M^{-1}E(U_1U_2S)-M^{-2}E(U_1S)E(U_2S).\\  $
	\subsection{3rd derivatives}
	\begin{align*}\frac{\partial^3 K(t_1,t_2)}{\partial t_{1}^3}&=M^{-1}\frac{\partial E(U_1^2S)}{\partial t_1}+E(U_1^2S)\frac{\partial M^{-1}}{\partial t_1}-M^{-2}\frac{\partial E(U_1S)E(U_1S)}{\partial t_1}-E(U_1S)E(U_1S)\frac{\partial M^{-2}}{\partial t_1}\\ &=M^{-1}E(U_1^3S)-M^{-2}E(U_1^2S)E(U_1S)-2M^{-2}E(U_1S)E(U_1^2S)+2M^{-3}(E(U_1S))^3,
	\end{align*}
	\begin{align*}
	\frac{\partial^3 K(t_1,t_2)}{\partial t_{2}^3}&=M^{-1}\frac{\partial E(U_2^2S)}{\partial t_2}+E(U_2^2S)\frac{\partial M^{-1}}{\partial t_2}-M^{-2}\frac{\partial E(U_2S)E(U_2S)}{\partial t_2}-E(U_2S)E(U_2S)\frac{\partial M^{-2}}{\partial t_2}\\  &=M^{-1}E(U_2^3S)-M^{-2}E(U_2^2S)E(U_2S)-2M^{-2}E(U_2S)E(U_2^2S)+2M^{-3}(E(U_2S))^3, 
	\end{align*}
	\begin{align*}
	\frac{\partial^3 K(t_1,t_2)}{\partial t_{1}^2 \partial t_2}&=M^{-1}\frac{\partial E(U_1^2S)}{\partial t_2}+E(U_1^2S)\frac{\partial M^{-1}}{\partial t_2}-M^{-2}\frac{\partial E(U_1S)E(U_1S)}{\partial t_2}-E(U_1S)E(U_1S)\frac{\partial M^{-2}}{\partial t_2}\\& =M^{-1}E(U_1^2U_2S)-M^{-2}E(U_1^2S)E(U_2S)-2M^{-2}E(U_1U_2S)E(U_1S)+2M^{-3}(E(U_1S))^2E(U_2S),
	\end{align*}
	\begin{align*}
	\frac{\partial^3 K(t_1,t_2)}{\partial t_{2}^2 \partial t_1}&=M^{-1}\frac{\partial E(U_2^2S)}{\partial t_1}+E(U_2^2S)\frac{\partial M^{-1}}{\partial t_1}-M^{-2}\frac{\partial E(U_2S)E(U_2S)}{\partial t_1}-E(U_2S)E(U_2S)\frac{\partial M^{-2}}{\partial t_1}\\& =M^{-1}E(U_2^2U_1S)-M^{-2}E(U_2^2S)E(U_1S)-2M^{-2}E(U_1U_2S)E(U_2S)+2M^{-3}(E(U_2S))^2E(U_1S). \end{align*}
	\subsection{4th derivatives}
	\begin{align*}
	\frac{\partial^4 K(t_1,t_2)}{\partial t_{1}^4}=&M^{-1}\frac{\partial E(U_1^3S)}{\partial t_1}+E(U_1^3S)\frac{\partial M^{-1}}{\partial t_1}-M^{-2}\frac{\partial E(U_1^2S)E(U_1S)}{\partial t_1}-E(U_1^2S)E(U_1S)\frac{\partial M^{-2}}{\partial t_1} \\&-2M^{-2}\frac{\partial E(U_1S)E(U_1^2S)}{\partial t_1}-2E(U_1S)E(U_1^2S)\frac{\partial M^{-2}}{\partial t_1}+2M^{-3}\frac{\partial (E(U_1S))^3}{\partial t_1}\\&+2(E(U_1S))^3\frac{\partial M^{-3}}{\partial t_1}\\&=M^{-1}E(U^4_1S)-M^{-2}E(U_1S)E(U_1^3S)-M^{-2}[E(U_1^3S)E(U_1S)+(E(U_1^2S))^2]\\&+2M^{-3}(E(U_1S))^2E(U_1^2S)-2M^{-2}[(E(U_1^2S)^2)+E(U_1S)E(U_1^3S)]\\&+4M^{-3}(E(U_1S))^2E(U_1^2S)+2M^{-3}[3(E(U_1S))^2E(U_1^2S)]-6M^{-4}(E(U_1S))^4,
	\end{align*}
	\begin{align*}
	\frac{\partial^4 K(t_1,t_2)}{\partial t_{1}^3\partial t_2}=&M^{-1}\frac{\partial E(U_1^3S)}{\partial t_2}+E(U_1^3S)\frac{\partial M^{-1}}{\partial t_2}-M^{-2}\frac{\partial E(U_1^2S)E(U_1S)}{\partial t_2}-E(U_1^2S)E(U_1S)\frac{\partial M^{-2}}{\partial t_2} \\&-2M^{-2}\frac{\partial E(U_1S)E(U_1^2S)}{\partial t_2}-2E(U_1S)E(U_1^2S)\frac{\partial M^{-2}}{\partial t_2}+2M^{-3}\frac{\partial (E(U_1S))^3}{\partial t_2}\\&+2(E(U_1S))^3\frac{\partial M^{-3}}{\partial t_2}
	\\&=M^{-1}E(U_1^3U_2S)-M^{-2}E(U_2S)E(U_1^3S)-M^{-2}[E(U_1^2U_2S)E(U_1S)+E(U_1U_2S)E(U_1^2S)]\\&+2M^{-3}E(U_2S)E(U_1S)E(U_1^2S)-2M^{-2}[E(U_1U_2S)E(U_1^2S)+E(U_1^2U_2S)E(U_1S)]\\&+4M^{-3}E(U_1S)E(U_2S)E(U_1^2S)+6M^{-3}(E(U_1S))^2E(U_1U_2S)-6M^{-4}(E(U_1S))^3E(U_2S),
	\end{align*}
	\begin{align*}
	\frac{\partial ^4K(t_1,t_2)}{\partial t_1^2t_2^2}=&M^{-1}\frac{\partial E(U_1U_2^2S)}{\partial t_1}+E(U_1U_2^2S)\frac{\partial M^{-1}}{\partial t_1}-M^{-2}\frac{\partial E(U_2^2S)E(U_1S)}{\partial t_1}-E(U_2^2S)E(U_1S)\frac{\partial M^{-2}}{\partial t_1}\\&-2M^{-2}\frac{\partial E(U_1U_2S)E(U_2S)}{\partial t_1}-2E(U_1U_2S)E(U_2S)\frac{\partial M^{-2}}{\partial t_1}+2M^{-3}\frac{\partial E(U_1S)(E(U_2S))^2}{\partial t_1}\\&+2E(U_1S)(E(U_2S))^2\frac{\partial M^{-3}}{\partial t_1}\\&=M^{-1}E(U_1^2U_2^2S)-M^{-2}E(U_1U_2^2S)E(U_1S)-M^{-2}[E(U_1U_2^2S)E(U_1S)+E(U_1^2S)E(U_2^2S)]\\&+2M^{-3}E(U_1S)E(U_1S)E(U_2^2S)-2M^{-2}[E(U_1^2U_2S)E(U_2S)+E(U_1U_2S)E(U_1U_2S)]\\&+4M^{-3}E(U_1S)E(U_2S)E(U_1U_2S)+2M^{-3}[E(U_1^2S)(E(U_2S))^2+2E(U_2S)E(U_1U_2S)E(U_1S)]\\&-6M^{-4}(E(U_1S))^2(E(U_2S))^2,
	\end{align*}
	\begin{align*}
	\frac{\partial^4 K(t_1,t_2)}{\partial t_{2}^3\partial t_1}&=M^{-1}\frac{\partial E(U_2^3S)}{\partial t_1}+E(U_2^3S)\frac{\partial M^{-1}}{\partial t_1}-M^{-2}\frac{\partial E(U_2^2S)E(U_2S)}{\partial t_1}-E(U_2^2S)E(U_2S)\frac{\partial M^{-2}}{\partial t_1} \\&-2M^{-2}\frac{\partial E(U_2S)E(U_2^2S)}{\partial t_1}-2E(U_2S)E(U_2^2S)\frac{\partial M^{-2}}{\partial t_1}+2M^{-3}\frac{\partial (E(U_2S))^3}{\partial t_1}\\&+2(E(U_2S))^3\frac{\partial M^{-3}}{\partial t_1}
	=M^{-1}E(U_2^3U_1S)-M^{-2}E(U_1S)E(U_2^3S)-M^{-2}[E(U_2^2U_1S)E(U_2S)\\&+E(U_2U_1S)E(U_2^2S)]+2M^{-3}E(U_1S)E(U_2S)E(U_2^2S)-2M^{-2}[E(U_1U_2S)E(U_2^2S)\\&+E(U_2^2U_1S)E(U_2S)]+4M^{-3}E(U_2S)E(U_1S)E(U_2^2S)+6M^{-3}(E(U_2S))^2E(U_2U_1S)\\&-6M^{-4}(E(U_2S))^3E(U_1S),
	\end{align*}
	\begin{align*}
	\frac{\partial^4 K(t_1,t_2)}{\partial t_{2}^4}&=M^{-1}\frac{\partial E(U_2^3S)}{\partial t_2}+E(U_2^3S)\frac{\partial M^{-1}}{\partial t_2}-M^{-2}\frac{\partial E(U_2^2S)E(U_2S)}{\partial t_2}-E(U_2^2S)E(U_2S)\frac{\partial M^{-2}}{\partial t_2}\\&-2M^{-2}\frac{\partial E(U_2S)E(U_2^2S)}{\partial t_2}-2E(U_2S)E(U_2^2S)\frac{\partial M^{-2}}{\partial t_2}\\&+2M^{-3}\frac{\partial (E(U_2S))^3}{\partial t_2}+2(E(U_2S))^3\frac{\partial M^{-3}}{\partial t_2}\\&=M^{-1}E(U^4_2S)-M^{-2}E(U_2S)E(U_2^3S)-M^{-2}[E(U_2^3S)E(U_2S)+(E(U_2^2S))^2]\\&+2M^{-3}(E(U_2S))^2E(U_2^2S)-2M^{-2}[(E(U_2^2S)^2)+E(U_2S)E(U_2^3S)]\\&+4M^{-3}(E(U_2S))^2E(U_2^2S)+2M^{-3}[3(E(U_2S))^2E(U_2^2S)]-6M^{-4}(E(U_2S))^4.
	\end{align*}
	\subsection{Cumulants}
	$k_1, k_2, k_3$ and $k_4$ represent the first, second, third ,fourth cumulant respectively.
	\begin{align*}
	\frac{\partial K(t_1,t_2)}{\partial t_1}|_{t_1=0,t_2=0}=E(U_1),
	\end{align*}
	\begin{align*}
	\frac{\partial K(t_1,t_2)}{\partial t_2}|_{t_1=0,t_2=0}=E(U_2),
	\end{align*}
	\begin{align*}
	k_1=E(U_1)+E(U_2).\\
	\end{align*}
	The first cumulant is done.
	\begin{align*}
	\frac{\partial^2 K(t_1,t_2)}{\partial t_1^2}|_{t_1=0,t_2=0}=E(U_1^2)-(E(U_1))^2,
	\end{align*}
	\begin{align*}
	\frac{\partial^2 K(t_1,t_2)}{\partial t_1 \partial t_2}|_{t_1=0,t_2=0}=E(U_1U_2)-E(U_1)E(U_2),
	\end{align*}
	\begin{align*}
	\frac{\partial^2 K(t_1,t_2)}{\partial t_2^2}|_{t_1=0,t_2=0}=E(U_2^2)-(E(U_2))^2,
	\end{align*}
	\begin{align*}
	k_2=E(U_1^2)-(E(U_1))^2+E(U_1U_2)-E(U_1)E(U_2)+E(U_2^2)-(E(U_2))^2.
	\end{align*}
	The second cumulant is done.
	\begin{align*}
	\frac{\partial^3 K(t_1,t_2)}{\partial t_1^3}|_{t_1=0,t_2=0}=E(U_1^3)-3E(U_1^2)E(U_1)+2(E(U_1))^3,
	\end{align*}
	\begin{align*}
	\frac{\partial^3 K(t_1,t_2)}{\partial t_2^3}|_{t_1=0,t_2=0}=E(U_2^3)-3E(U_2^2)E(U_2)+2(E(U_2))^3,
	\end{align*}
	\begin{align*}
	\frac{\partial^3 K(t_1,t_2)}{\partial t_1^2 \partial t_2}|_{t_1=0,t_2=0}=E(U_1U_2^2)-E(U_2^2)E(U_1)-2E(U_1U_2)E(U_2)+2(E(U_2))^2E(U_1),
	\end{align*}
	\begin{align*}
	\frac{\partial^3 K(t_1,t_2)}{\partial t_1 \partial t_2^2}|_{t_1=0,t_2=0}=E(U_1^2U_2)-E(U_1^2)E(U_2)-2E(U_1U_2)E(U_1)+2(E(U_1))^2E(U_2).
	\end{align*}
	\begin{align*}
	k_3=E(U_1^3)+E(U_2^3)+E(U_1U_2^2)+E(U_1^2U_2)-3E(U_1^2)E(U_1)-3E(U_2^2)E(U_2)-E(U_1)E(U_2^2)-E(U_1^2)E(U_2)\\-2E(U_1)E(U_1U_2)-2E(U_2)E(U_1U_2)+2E(U_1)(E(U_2))^2+2(E(U_1))^2E(U_2)+2(E(U_1))^3+2(E(U_2))^3.
	\end{align*}
	The third cumulant is done.
	\begin{align*}
	\frac{\partial^4 K(t_1,t_2)}{\partial t_1^4}|_{t_1=0,t_2=0}=E(U_1^4)-E(U_1)E(U_1^3)-E(U_1^3)E(U_1)-(E(U_1^2))^2+2(E(U_1))^2E(U_1^2)\\-2[(E(U_1^2))^2+E(U_1)E(U_1^3)]+4(E(U_1))^2E(U_1^2)+6(E(U_1))^2E(U_1^2)-6(E(U_1))^4,
	\end{align*}
	\begin{align*}
	\frac{\partial^4 K(t_1,t_2)}{\partial t_2^4}|_{t_1=0,t_2=0}=E(U_2^4)-E(U_2)E(U_2^3)-E(U_2^3)E(U_2)-(E(U_2^2))^2+2(E(U_2))^2E(U_2^2)\\-2[(E(U_2^2))^2+E(U_2)E(U_2^3)]+4(E(U_2))^2E(U_2^2)+6(E(U_2))^2E(U_2^2)-6(E(U_2))^4,
	\end{align*}
	\begin{align*}
	\frac{\partial^4 K(t_1,t_2)}{\partial t_1^3 \partial t_2}|_{t_1=0,t_2=0}=E(U_1^3U_2)-E(U_1^3)E(U_2)-E(U_1)E(U_1^2U_2)-E(U_1^2)E(U_1U_2)+2E(U_1)E(U_2)E(U_1^2)\\-2E(U_1^2)E(U_1U_2)-2E(U_1)E(U_1^2U_2)+4E(U_1)E(U_2)E(U_1^2)+6(E(U_1))^2E(U_1U_2)-6(E(U_1))^3E(U_2),
	\end{align*}
	\begin{align*}
	\frac{\partial^4 K(t_1,t_2)}{\partial t_1 \partial t_2^3}|_{t_1=0,t_2=0}=E(U_1U_2^3)-E(U_1)E(U_2^3)-E(U_2)E(U_2^2U_1)-E(U_2^2)E(U_1U_2)+2E(U_1)E(U_2)E(U_2^2)\\-2E(U_2^2)E(U_1U_2)-2E(U_2)E(U_1U_2^2)+4E(U_1)E(U_2)E(U_2^2)+6(E(U_2))^2E(U_1U_2)-6(E(U_2))^3E(U_1),
	\end{align*}
	\begin{align*}
	\frac{\partial^4 K(t_1,t_2)}{\partial t_1^2 \partial t_2^2}|_{t_1=0,t_2=0}=E(U_1^2U_2^2)-2E(U_1)E(U_1U_2^2)-E(U_1^2)E(U_2^2)+2(E(U_1))^2E(U_2^2)-2E(U_2)E(U_1^2U_2)\\-2E((U_1U_2))^2+4E(U_1)E(U_2)E(U_1U_2)+2E(U_1^2)(E(U_2))^2+4E(U_1)E(U_2)E(U_1U_2)-6(E(U_1))^2(E(U_2))^2,
	\end{align*}
	\begin{align*}
	k_4=&E(U_1^4)-E(U_1)E(U_1^3)-E(U_1^3)E(U_1)-(E(U_1^2))^2+2(E(U_1))^2E(U_1^2)-2[(E(U_1^2))^2+E(U_1)E(U_1^3)]\\&+4(E(U_1))^2E(U_1^2)+6(E(U_1))^2E(U_1^2)-6(E(U_1))^4+E(U_2^4)-E(U_2)E(U_2^3)-E(U_2^3)E(U_2)-(E(U_2^2))^2\\&+2(E(U_2))^2E(U_2^2)-2[(E(U_2^2))^2+E(U_2)E(U_2^3)]+4(E(U_2))^2E(U_2^2)+6(E(U_2))^2E(U_2^2)-6(E(U_2))^4\\&+E(U_1^3U_2)-E(U_1^3)E(U_2)-E(U_1)E(U_1^2U_2)-E(U_1^2)E(U_1U_2)+2E(U_1)E(U_2)E(U_1^2)-\\&2E(U_1^2)E(U_1U_2)-2E(U_1)E(U_1^2U_2)+4E(U_1)E(U_2)E(U_1^2)+6(E(U_1))^2E(U_1U_2)-6(E(U_1))^3E(U_2)\\&+E(U_1U_2^3)-E(U_1)E(U_2^3)-E(U_2)E(U_2^2U_1)-E(U_2^2)E(U_1U_2)+2E(U_1)E(U_2)E(U_2^2)-\\&2E(U_2^2)E(U_1U_2)-2E(U_2)E(U_1U_2^2)+4E(U_1)E(U_2)E(U_2^2)+6(E(U_2))^2E(U_1U_2)-6(E(U_2))^3E(U_1)\\&E(U_1^2U_2^2)-2E(U_1)E(U_1U_2^2)-E(U_1^2)E(U_2^2)+2(E(U_1))^2E(U_2^2)-2E(U_2)E(U_1^2U_2)-2E((U_1U_2))^2\\&+4E(U_1)E(U_2)E(U_1U_2)+2E(U_1^2)(E(U_2))^2+4E(U_1)E(U_2)E(U_1U_2)-6(E(U_1))^2(E(U_2))^2\\=&E(U_1^4)-4E(U_1)E(U_1^3)-3(E(U_1^2))^2+12(E(U_1))^2E(U_1^2)-6(E(U_1))^4\\&+E(U_2^4)-4E(U_2)E(U_2^3)-3(E(U_2^2))^2+12(E(U_2))^2E(U_2^2)-6(E(U_2))^4\\&+E(U_1^3U_2)-E(U_1^3)E(U_2)-3E(U_1)E(U_1^2U_2)-3E(U_1^2)E(U_1U_2)-6E(U_1)E(U_2)E(U_1^2)\\&+6E(U_1^2)E(U_1U_2)-6(E(U_1))^3E(U_2)\\&+E(U_1U_2^3)-E(U_1)E(U_2^3)-3E(U_2)E(U_2^2U_1)-3E(U_2^2)E(U_1U_2)+6E(U_1)E(U_2)E(U_2^2)\\&+6E(U_2^2)E(U_1U_2)-6(E(U_2))^3E(U_1)\\&E(U_1^2U_2^2)-2E(U_1)E(U_1U_2^2)-E(U_1^2)E(U_2^2)+2(E(U_1))^2E(U_2^2)-2E(U_2)E(U_1^2U_2)-2E((U_1U_2))^2\\&+8E(U_1)E(U_2)E(U_1U_2)+2E(U_1^2)(E(U_2))^2-6(E(U_1))^2(E(U_2))^2.
	\end{align*}
	The fourth cumulant is done.


\begin{thebibliography}{9}

	\bibitem{Test Variables}
	H. B. Mann and D. R. Whitney. On a test whether one of two random variables is stochastically larger than the other. 
	\textit{The Annals of Mathematical Statistics} 1850-60, 1947
		\bibitem{Spurrier}
	John D. Spurrier and John E. Hewett. Two-Stage Wilcoxon Tests of Hypotheses.
	\textit{Journal of the American Statistical Association} Vol. 71, No. 356 (Dec., 1976), pp. 982-987
		\bibitem{kolassa} 
	J. E. Kolassa. A comparison of size and power calculations for the Wilcoxon Statistic for ordered categorical data.
	\textit{Statistics in Medicine}. 
	VOL. 14, 1577-1581, 1995.	
\end{thebibliography}
\end{document}